\documentclass[
reprint,
groupedaddress,
amsmath,amssymb,
aps,prl,
twocolumn,
showpacs,
]{revtex4-1}

\usepackage{tikz}
\usepackage{graphicx}
\usepackage{dcolumn}
\usepackage{bm}
\usepackage{braket}
\usepackage{hyperref}
\usepackage[caption=false]{subfig}
\usepackage{amsmath}
\usepackage{mathtools} 
\usepackage{nicematrix} 
\usepackage{graphicx}
\usepackage{fancyhdr}
\usepackage{float}
\usepackage{siunitx}
\usepackage{subfig}
\usepackage{xcolor}
\usepackage{xspace}
\usepackage{eucal}
\usepackage{multirow}
\usepackage[utf8]{inputenc}
\usepackage{pgfplots}
\pgfplotsset{compat=1.14}
\begin{document}

\DeclareSIUnit{\belmilliwatt}{Bm}
\DeclareSIUnit{\dBm}{\deci\belmilliwatt}
\DeclareSIUnit{\gammas}{$\Gamma$}
\DeclareSIUnit{\gauss}{G}

\newcommand{\update}[1]{\textbf{\color{red}UPDATE:~#1}\xspace}
\newcommand{\strike}[1]{\color{red}\sout{#1}\xspace}
\newcommand{\new}[1]{{\color{green}#1}\xspace}
\sisetup{detect-weight=true, detect-family=true}

\definecolor{myblue}{RGB}{0,58,109}
\definecolor{myred}{RGB}{204,0,1}
\definecolor{mygreen}{RGB}{11,102,35}

\title{
Multi-axis inertial sensing with 2D matter-wave arrays
}
\author{K.~Stolzenberg}
\author{C.~Struckmann}
\author{S.~Bode}
\author{R.~Li}
\author{A.~Herbst}
\author{V.~Vollenkemper}
\author{D.~Thomas}
\author{A.~Rajagopalan}
\author{E.~M.~Rasel}
\author{N.~Gaaloul}
\author{D.~Schlippert}\email{schlippert@iqo.uni-hannover.de}
\affiliation{Leibniz Universit\"at Hannover, Institut f\"ur Quantenoptik,\\ Welfengarten 1, 30167 Hannover, Germany}

\date{\today}

\begin{abstract}
Atom interferometery is an exquisite measurement technique sensitive to inertial forces. However, it is commonly limited to a single sensitive axis, allowing high-precision multi-dimensional sensing only through subsequent or post-corrected measurements.
We report on a novel method for multi-axis inertial sensing based on the correlation of simultaneous light-pulse atom interferometers in 2D array arrangements of Bose-Einstein Condensates (BEC). 
Deploying a scalable $3\times 3$ BEC array spanning \SI{1.6}{\milli\meter\squared} created using time-averaged optical potentials, we perform measurements of linear acceleration induced by gravity and simultaneously demonstrate sensitivity to angular velocity and acceleration of a rotating reference mirror, as well as gravity gradients and higher-order derivatives.
Our work enables simple, high-precision multi-axis inertial sensing compatible with high rotation rates, e.g., for inertial navigation in dynamic environments.
We finally envision further applications of our method, e.g., 3D in-situ measurements and reconstruction of laser beam intensities and wave fronts.
\end{abstract}

\maketitle
\textbf{Introduction} -
Quantum sensors based on light-pulse atom interferometry~\cite{Kasevich91PRL} provide a highly sensitive and long-term stable measurement tool for inertial forces~\cite{Geiger2011}, such as linear accelerations~\cite{Peters_2001_gravity}, gravity gradients and curvature~\cite{mcguirk_sensitive_2002,rosi_measurement_2015,Asenbaum17PRL}, or rotation rates \cite{Boshier_gyro,gautier_accurate_2022,berg_composite-light-pulse_2015,stockton_absolute_2011}.
However, conventional atom interferometers feature only one sensitive axis, yielding intertwined information about one acceleration and one rotation component. To resolve individual inertial quantities, correlation with another simultaneous interferometer or external inertial sensor is necessary~\cite{decastanet2024atom,canuel_six-axis_2006,barrett_multidimensional_2019,gersemann_differential_2020,ledesma2024vectoratomaccelerometryoptical}. 
\begin{figure}[h!]
    \begin{center}
    \resizebox{0.99\columnwidth}{!}{\includegraphics[width=1\textwidth]{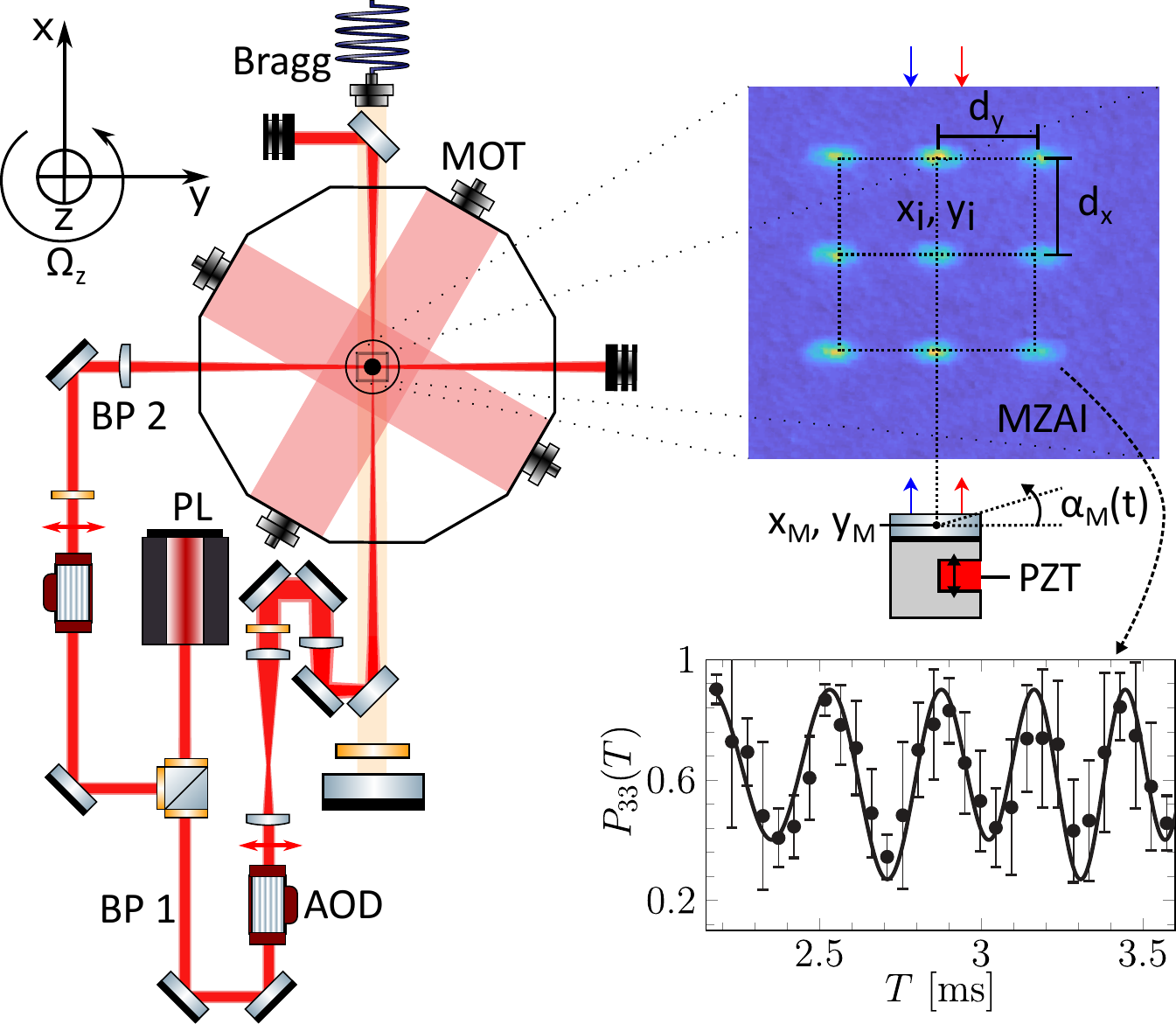}}
     \caption{\textbf{Experimental setup:} Left: Top view highlighting the optical dipole trap (ODT) setup, with $z$ being aligned in direction of gravity $g$. The two dipole trap beams (BP1/2) intersect at an angle of approximately $\SI{90}{\degree}$. The two 2D acousto-optical deflectors (AOD) in the beam paths allow for dynamically deflecting the laser beams, establishing a 2D array of potential wells. The Bragg beam is aligned with BP1 and split at a dichroic mirror. Top right: Absorption image of a $3\times3$ array and illustration of the piezo (PZT), allowing to tilt the reference mirror by an angle $\alpha_M(t)$. Bottom right: Typical fringe measurement (dots) by scanning the evolution time $T$ for a single ensemble. Due to induced tilts of the reference mirror and free fall through the Bragg beam, the beam splitter's efficiency is not ideal, resulting in the beating of the fringe contrast as predicted by a double-Bragg-diffraction ab-inito theory~\cite{Li2024PRR}.}  
    \label{fig:Setup}
    \end{center}
\end{figure}
Point-source atom interferometers expand the dimensionality of the inertial observables by exploiting the 2D surface of the thermal ensemble or Bose-Einstein condensate (BEC), enabling the simultaneous measurement of two rotation axes~\cite{Dickerson13PRL,Chen2019PSI}. However, they require clouds with large spatial extent and are thus limited to operate with thermal ensembles in compact setups.
The limited sensitivity scaling and dynamic range can be enhanced by considering BECs, providing a coherent source over short and long time scales~\cite{Deppner21PRL}.
1D arrays of BECs have been introduced in magnetic lattices~\cite{Jose_2014_magnetic} and optical tweezers~\cite{Gosar_2022_tweezers} for atom interferometry, e.g., by exploiting the tunneling effect~\cite{nemirovsky2023atomic}.
Very recently, 2D arrays of BECs have been used to study collapse dynamics~\cite{huang2024two}.
\\
\begin{figure*}[t!]
    \begin{center}
    \includegraphics[width=.995\textwidth]{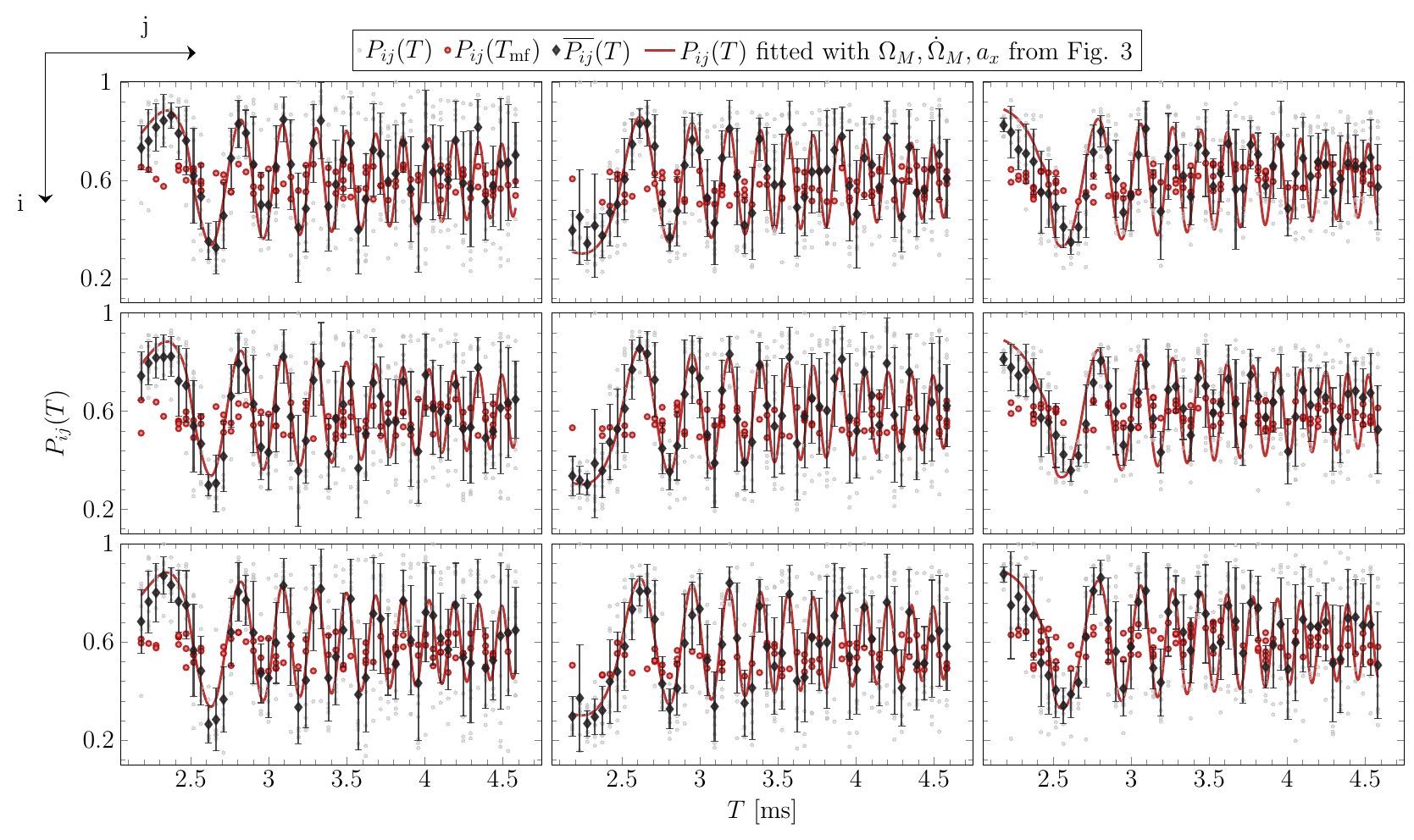}
    \caption{\textbf{Evolution time scans of a $3\times3$-array with the induced tilting of the retro reflection mirror:} Fringes obtained for the 9 simultaneous atom interferometers for $T=2.182-4.582$ \SI{}{\milli\second}. Light gray circles $P_{ij}(T)$ show the outcome of each of the 510 measurements per interferometer. The average of the data $\overline{P_{ij}}(T)$ for the 10 successive experiment runs are depicted with black diamonds. Red squares mark the mid-fringe data points $P_{ij}(T_{\mathrm{mf}})$. The inertial observables ($a_x, \Omega_M, \dot{\Omega}_M$) were differentially deduced (Fig. \ref{fig:phase-rot}) and fitted with an exponential contrast loss and offset phase shown as red line.}
    \label{fig:fringes}
    \end{center}
\end{figure*}
In this Letter, we report on the correlation of multiple light-pulse atom interferometers generated from 2D BEC arrays allowing for a simultaneous measurement of several inertial observables.
We refer to this approach as PIXL (Parallelized Interferometers for XLerometry).
The 2D BEC array with scalable dimensionality and spacing of up to $3\times3$ across \SI{1.6}{\milli\meter\squared} (Fig.~\ref{fig:Setup}) is created using the exquisite trap control granted by time-averaged optical potentials~\cite{Roy2016PRA,Condon2019,Herbst2024highflux} in a crossed optical dipole trap (ODT), utilizing $^{87}$Rb atoms.
By using each array site as a source for a Mach-Zehnder type atom interferometer (MZAI) in which double-Bragg diffraction~\cite{Ahlers16PRL} coherently splits, redirects, and recombines matter waves, we realize nine strongly correlated 1D linear acceleration measurements.
Combinations of two or more 1D linear acceleration measurements separated by a scalable baseline give rise to common-mode noise rejection and sensitivity to gravity gradients and higher derivatives as well as to angular velocity and acceleration, similar to inertial-stabilized platforms~\cite{wanner_seismic_2012}, classical accelerometer arrays~\cite{Schuler1967} or as proposed for combining multiple multi-axis atom interferometers~\cite{Shettel2023}.\\
\textbf{Phase sensitivity} - \label{sec:phase_sensitivity}
The phase response of each of the single atom interferometers of the array is subject to all inertial forces acting on the reference mirror.
Rotational effects induce phase shifts scaling with the initial kinematics of the BECs with respect to the reference mirror and thus, are of special interest here.
Generally, considering all three possible axes, we can write the phase signal in terms of the linear acceleration vector $\mathbf{a}$, linear acceleration gradient tensor $\hat\Gamma$, angular velocity tensor $\hat\Omega$ and angular acceleration tensor $\dot{\hat\Omega}$,
\begin{widetext}
    \begin{equation}
    \mathbf{a} =
    \begin{pNiceMatrix}
        \color{myred} a_x \\
        a_y \\
        a_z \\
    \end{pNiceMatrix}, \ 
    \hat\Gamma = 
    \begin{pNiceMatrix}[first-row,last-col]
          \color{myred} \scriptstyle\Delta x & \color{myred} \scriptstyle\Delta y & \scriptstyle\Delta z & \\
          \color{myred} \tfrac{\partial a_x}{\partial x} & \color{myred} \tfrac{\partial a_x}{\partial y} & \tfrac{\partial a_x}{\partial z} \\
          \tfrac{\partial a_y}{\partial x} & \tfrac{\partial a_y}{\partial y} & \tfrac{\partial a_y}{\partial z} \\
          \tfrac{\partial a_z}{\partial x} & \tfrac{\partial a_z}{\partial y} & \tfrac{\partial a_z}{\partial z} \\
    \end{pNiceMatrix}, \ 
    (\hat\Omega)^2 = 
    \begin{pNiceMatrix}[first-row,last-col]
          \color{myred} \scriptstyle\Delta x & \color{myred} \scriptstyle\Delta y & \scriptstyle\Delta z & \\
          \color{myred} -\Omega_y^2 - \Omega_z^2 & \color{myred} \Omega_x\Omega_y & \Omega_x\Omega_z \\
          \Omega_x\Omega_y & -\Omega_x^2 - \Omega_z^2 & \Omega_y\Omega_z \\
          \Omega_x\Omega_z & \Omega_y\Omega_z & -\Omega_x^2 - \Omega_y^2 \\
    \end{pNiceMatrix}, \ 
    \dot{\hat\Omega} = 
    \begin{pNiceMatrix}[first-row,last-col]
          \color{myred} \scriptstyle\Delta x & \color{myred} \scriptstyle\Delta y & \scriptstyle\Delta z & \\
          \color{myred} 0 & \color{myred} -\dot\Omega_z & \dot\Omega_y & \color{myred} \scriptstyle k_{\mathrm{eff},x}\\
          \dot\Omega_z & 0 & -\dot\Omega_x & \scriptstyle k_{\mathrm{eff},y} \\
          -\dot\Omega_y & \dot\Omega_x & 0 & \scriptstyle k_{\mathrm{eff},z} \\
    \end{pNiceMatrix}.
    \label{eq:inertial-quantities}
\end{equation}    
\end{widetext}
For an arbitrarily oriented sensitive axis, defined by 
$\mathbf{k_\mathrm{eff}} = (k_{\mathrm{eff},x},k_{\mathrm{eff},y},k_{\mathrm{eff},z})^T$,
we find with the approach demonstrated in Refs.~\cite{beaufils2023rotation,Struckmann2024},
\begin{equation}
    \phi_{ijk} = 2 T^2 \mathbf{k_\mathrm{eff}} \cdot [\mathbf{a} + (\hat\Gamma - \dot{\hat\Omega} - (\hat\Omega)^2)\cdot(\mathbf{r}_{ijk}-\mathbf{r}_M)],
\end{equation}
where rotations of $\hat\Gamma$ and initial velocities were neglected. 
Note that we find the standard Euler and centrifugal acceleration caused by the angular acceleration and velocity, respectively.
The indices $i,j,k\in\mathbb N$ label the position in the 3D BEC array. $\mathbf{r}_{ijk}=(x_{ijk},y_{ijk},z_{ijk})^T$ and $\mathbf{r}_M = (x_M, y_M, z_M)^T$ denote the initial positions of the BECs and the mirror's rotation center, respectively.
The various components of the tensors are directly related to the experimental setup required for their measurement. Row-wise components can be measured by choosing different sensitive axes ($k_{\mathrm{eff},x}$, $k_{\mathrm{eff},y}$, $k_{\mathrm{eff},z}$), whereas measuring column-wise components relies on the differential positions available from the BEC array ($\Delta x$, $\Delta y$, $\Delta z$). Notably, a conventional atom interferometer with a stationary initial BEC is only able to detect the diagonal elements. Utilizing the spatial extent of a BEC is possible but offers very limited sensitivity due to the relatively small area covered.
\\
The inertial quantities  under consideration here are marked in red in eq.~\eqref{eq:inertial-quantities}. Our current PIXL setup comprises an interferometer beam aligned with the $x$-axis ($k_{\mathrm{eff},x}$) and a 2D BEC array oriented in the $x$-$y$-plane ($\Delta x$, $\Delta y$), allowing the measurement of $a_x$, $(\partial_x a_x + \Omega_y^2 + \Omega_z^2)\Delta x$ and $(\partial_y a_x - \Omega_x\Omega_y + \dot\Omega_z)\Delta y$.
Here, we consider rotations with constant angular acceleration $\dot{\Omega}_M$ around the $z$-axis (Fig.~\ref{fig:Setup}), $\alpha_M(t) = \alpha_{M,0} + \Omega_M (t-T) + \tfrac{1}{2}\dot{\Omega}_M(t-T)^2$, where $\Omega_M$ is the angular velocity, sampled around the temporal center of the atom interferometer sequence corresponding to the interrogation time $T$. Furthermore, we align the sensitive axis perpendicular to the gravitational acceleration $g$ up to a slight tilt $\theta_{g}$, inducing a small linear acceleration $a_{x,ij} = g \sin(\theta_{g})$.
The leading order atom interferometer phase signal is 
\begin{align}
     \phi_{ij} = 2 k_\mathrm{eff} T^2 [
     &a_{x,ij} + (y_{ij}-y_M)\dot{\Omega}_M \notag\\
     & + 2(x_{ij}-x_M)(\Omega_M^2 + \tfrac{7}{2}\dot{\Omega}_M^2 T^2)],
\label{eq:Sensitivity_Array}
\end{align}
including a higher order term accounting for high angular accelerations. 
We further neglected phase shifts induced by gravity gradients and Earth's rotation rate due to the limited interrogation times $T<\SI{5}{ms}$. 
As the rotation rate appears squared in the phase, we are restricted to measuring the absolute value of the rotation rate, $|\Omega_M|$. 
In a full 3D PIXL setup, the sign ambiguity can be resolved by considering the relationships between $\Omega$ and $\dot\Omega$~\cite{williams2013minimal}. Alternatively, a linear relation in the phase could be achieved by considering correlations between two BECs initialized with different velocities (see eq.~\ref{eq:appendix-ai-phase-full}). For simplicity, we will omit the absolute value notation in the following discussion. 
\\
\begin{figure}
    \centering
    \includegraphics[width=\linewidth]{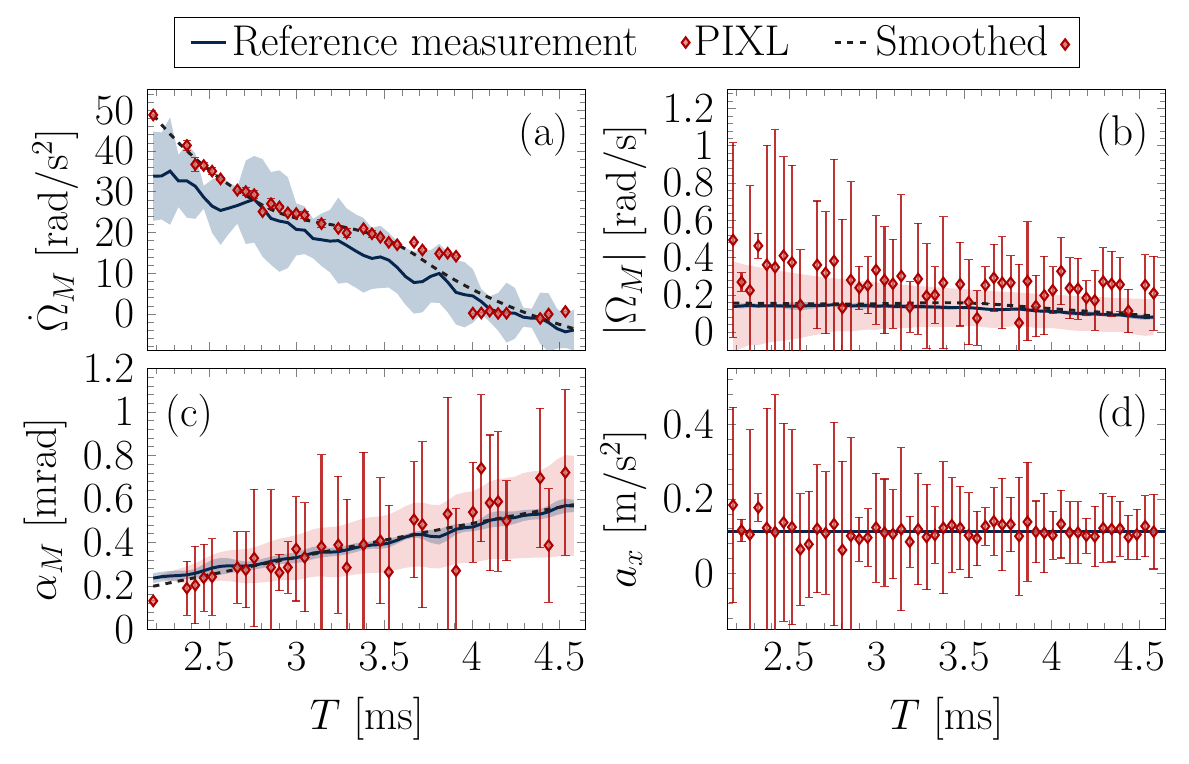}
    \caption{\textbf{Result of fitting the relative populations compared to a direct measurement.}
    By taking the column-wise and row-wise differences of the phases at the mid-fringe, the angular acceleration $\dot\Omega_M$ and angular velocity $\Omega_M$ are obtained in (a) and (b), respectively. 
    Further, this allows the reconstruction of the mirror's angle $\alpha_M$ up to a constant offset in (c). 
    Fitting an angle $\alpha_M$ defined by a spline interpolation between $5$ equidistant points to match the measured $\Omega_M$ and $\dot\Omega_M$ smooths the measurement points (dashed lines).
    The results obtained with PIXL (red diamonds) are compared to a direct measurement of the mirror's rotation (solid blue line).
    The quantum projection noise limit is shaded in light red around the direct measurements.
    The uncertainty, which corresponds to the reproducibility of the mirror rotation from shot to shot, is shaded blue.
    Using these rotational observables, the linear acceleration can be isolated from the phase, eq.~\eqref{eq:Sensitivity_Array}. 
    Note that this requires a preceding calibration of the mirror's rotation center with respect to the BEC array's center, $x_M = 0.250(5)\SI{}{\meter}$ and $y_M=\SI{2.0(4)}{\milli\meter}$. The remaining linear acceleration obtained from the fringes (red diamonds), Fig.~\ref{fig:fringes}, is depicted in (d) together with the average acceleration measured without rotations (solid blue line).
    For each $T$, an inconsistent number of mid-fringe positions in each BEC was found leading to high fluctuations of the uncertainties of the atom interferometric measurement.}
    \label{fig:phase-rot}
\end{figure}
\textbf{Rotation measurement} -
For the demonstration and characterization of PIXL, we initialize a $3 \times 3$ array with spatial separation of $(d_{x}, d_{y}) = (1.0,0.435)$~mm, negligible initial velocity and in the $\ket{\mathrm{F}=1,\mathrm{m_F}=0}$ state. 
A detailed description of the production, state preparation, and transport of the BEC array, as well as the interferometry setup, is provided in the supplemental material (see also references 
~\cite{Torrontegui2011,Giese13PRA,küber2016experimental,Noh2011MTS,Parker2016PRA, Bonnin13PRA} therein) attached after the main text. 
We subsequently release the BECs from the crossed ODT and apply the first interferometer pulse after \SI{1}{\milli\second} of free fall.
We scan the interrogation time $T$ from \SI{2.182}{ms} to \SI{4.582}{ms} in 50 equidistant steps to obtain the atom interferometer phase $\phi_{ij}(T)$ in post processing (cf. eq.~\eqref{eq:Sensitivity_Array}).
The experiment was calibrated and analyzed alongside a sophisticated ab-inito theoretical model, which was further simplified to highlight key figures of merit relevant to this study~\cite{Li2024PRR}.
The signal under consideration is the normalized population, $P_{ij}=N_{ij,\pm\hbar k_\mathrm{eff}}/N_{ij,\mathrm{tot}}$, of the $\pm\hbar k_\mathrm{eff}$ states which is of the form, 
\begin{align}
    P_{ij}(T)=P_{ij,0}+\tfrac{1}{2}C_{ij}\times\sin(\phi_{ij}(T)+\phi_0),
    \label{eqn:signal}
\end{align}
where $P_{ij,0}$ denotes a population offset, $C_{ij}$ the contrast of the interferometer and $\phi_0$ a constant phase offset (cf. Fig.~\ref{fig:fringes}).
We furthermore induce a site-dependent phase shift by applying an angular velocity of $\Omega_M \sim \SI{0.1}{\radian\per\second}$ and angular acceleration of $\dot{\Omega}_M \sim \SI{20}{\radian\per\second\squared}$ to the retro-reflecting mirror throughout the interferometer sequence using a piezo tip-tilt stage (cf. Fig.~\ref{fig:phase-rot} (c)).
\\
We focus on interrogation times $T_\mathrm{mf}$ around the mid-fringe positions, $|P_{ij}(T_\mathrm{mf}) - P_{ij,0}|\leq 0.11\times C_{ij}$, where the relative population can be assumed to be directly proportional to the phase shift, $P_{ij}(T_\mathrm{mf}) \approx P_{ij,0} \pm \tfrac{1}{2}C_{ij} \times (\phi_{ij}(T_\mathrm{mf}) + \phi_0 + n_{ij}\pi)$, $n_{ij}\in\mathbb{Z}$. 
For these times $T_\mathrm{mf}$, the angular velocity and angular acceleration can be directly obtained from the row-wise and column-wise differences of the phases: $\Omega_M^2(T_\mathrm{mf}) = (\phi_{i+1,j}-\phi_{i,j})/(4k_\mathrm{eff}d_xT_\mathrm{mf}^2)$ and $\dot\Omega_M(T_\mathrm{mf}) = (\phi_{i,j+1}-\phi_{i,j})/(2k_\mathrm{eff}d_yT_\mathrm{mf}^2)$. 
The complete data processing procedure is detailed in the supplemental material~\cite{stolzenberg2024sup}.
For large $\dot\Omega_M$, the second order term $\propto \dot\Omega_M^2T^4$ needs to be taken into account and subtracted from the $\Omega_M^2$.
Notably, this is based on single measurements, utilizing common-mode noise rejection. Subsequently, $\Omega_M(t)$ and $\dot\Omega_M(T)$ are averaged over individual measurements.
Furthermore, we constrain the temporal change of the angular velocity, 
$\Omega_M^2(T) \approx (\alpha_M^2(0)-2\alpha_M^2(T)+\alpha_M^2(2T))/(2T^2)$ 
to be consistent with the angular acceleration, $\dot\Omega_M(T) \approx (\alpha_M(0)-2\alpha_M(T)+\alpha_M(2T))/T^2$, using the discrete samples of the mirror angle $\alpha_M(t)$ at times $0$, $T$ and $2T$.
To further smooth the measurement results, we fit $\alpha_M(t)$ using spline interpolation between $5$ equidistant points to match the measured $\Omega_M$ and $\dot\Omega_M$ at the corresponding times.
The time series of $\dot\Omega_M(T)$, $\Omega_M(T)$ and $\alpha_M(T)$ are summarized in Fig.~\ref{fig:phase-rot}.
\\
We observe good qualitative agreement between the results obtained from fitting the interferometer phases and a direct measurement of the mirror's rotation obtained from deflection measurements.
The extraction of the linear acceleration from the dataset requires an estimate of $x_M$ and $y_M$, achievable by performing preceding calibration experiments using known mirror rotations or an additional imaging system. 
Here, we measure $x_M = \SI{0.250(5)}{\meter}$ and calibrate $y_M=\SI{2.0(4)}{\milli\meter}$ using the classical measurement.
Similarly, the linear acceleration $a_{x,ij}$ is determined by subtracting the rotation-induced accelerations from the total phase. This phase is derived by considering small time windows of constant accelerations, which are swept over the signal (see Fig.~\ref{fig:fromfringetoscf} and Fig.~\ref{fig:columnwisephases} in supplemental material).
The resulting linear acceleration is shown in Fig.~\ref{fig:phase-rot}. We find good consistency with a preceding atom interferometric measurement of just the residual linear acceleration (see Fig.~\ref{fig:fringeswo} in supplemental material). These measurements additionally incorporate a beating pattern, resulting from an imperfect beam-splitting pulse, as predicted by the double-Bragg-diffraction ab-inito theory~\cite{Li2024PRR}.  Deviations from the preceding measurement can be explained by a slight tilt of the mirror's rotational axis with respect to gravity.
\\
Finally, using eq.~\eqref{eqn:signal} we convert the total phase shift $\phi_{ij}(T)$ to relative populations $P_{ij}(T)$, fitting the contrast $C_{ij}$, population offset $P_{ij,0}$ and show the results along side with the experimental data in Fig.~\ref{fig:fringes}.
\\
\textbf{Performance assessment} -
We assess PIXL's sensitivity two-fold. On one hand, the uncertainty of linear acceleration measurements relies on the analysis of the individual sensitivities per array site. 
On the other hand, gradiometric and gyroscopic measurements are based on differential signals between individual BECs and benefit from common-mode rejection, thus only involving position-dependent noise sources.
PIXL's fundamental sensitivity limit to an acceleration $a$ is given by $\delta a_{ij}^\mathrm{SN} = \delta\phi_{ij}^{\rm SN}/(2 k_\mathrm{eff}T^2)$ where $\delta \phi_{ij}^\mathrm{SN} = 1/(C_{ij}\sqrt{N_{ij}})$ is the quantum projection noise floor. 
Here, $C_{ij}$ denotes the contrast and $N_{ij}$ the atom number of the $(i,j)$ atom interferometer. For our experiment with $N_{ij}=10^4, C_{ij}=0.6$ and $2T=\SI{6}{\milli\second}$, we find $\delta a_{ij}^\mathrm{SN} = \SI{6e-5}{\meter\per\second\squared}$, while all measurements are dominated by vibrational noise estimated at a level of $\sim \SI{6}{\milli\meter\per\second\squared}$ using the post correction formalism~\cite{Cheinet2008,Richardson2019}. We estimate $\sim\SI{34}{\decibel}$ common-mode noise rejection ratio towards vibrations (see Fig.~\ref{fig:postcorrection} b) in supplemental material) 
so vibrational noise is suppressed in the differential measurement of linear acceleration gradients or rotation rates, but requires the precise knowledge of the BECs' differential initial kinematics. 
For convenience, we define the pair-wise sensitivity limit, $\delta\phi_{j}^\mathrm{SN}=\sqrt{\sum_{i}(\delta\phi_{ij}^\mathrm{SN})^2}$ with $i$ ($j$) labeling rows (columns).
The shot noise limited sensitivity for acceleration gradients $\gamma$ is given by $\delta \gamma_{j}^\mathrm{SN} = \delta\phi_{j}^\mathrm{SN}/(2 d_x k_\mathrm{eff} T^2) = \SI{8.1e-2}{\per\second\squared}$.
To first order in $T^2$, the angular velocity $\Omega$ and acceleration $\dot\Omega$ can be inferred by correlating phase measurements of BECs distributed along or transversely to the beam direction. 
The sensitivity limits are 
$\delta\Omega_{j}^\mathrm{SN} = \delta\phi_{j}^\mathrm{SN}/(8 d_x \Omega k_\mathrm{eff} T^2)$
and $\delta\dot\Omega_{i}^\mathrm{SN} = \delta\phi_{i}^\mathrm{SN}/(2 d_y k_\mathrm{eff} T^2)$, respectively. Here, we find $\delta\Omega_{j}^{\mathrm{SN}} = \SI{2e-2}{\radian\squared\per\second\squared}/\Omega$ and $\delta\dot\Omega^\mathrm{SN}_{i} = \SI{1.9e-1}{\radian\per\second\squared}$.
For the considered angular velocity of $\Omega\sim \SI{0.1}{\radian\per\second}$, this yields $\delta\Omega_{j}^\mathrm{SN} = \SI{203}{\milli\radian\per\second}$.
Since the differential measurements are depending on the relative initial positions and velocities of the ensembles of the array, we analyzed the positions of the initial state of each interferometer and estimated positional uncertainties of $\sigma_{d_x} = \SI{8}{\micro\meter}$ and $\sigma_{d_y} = \SI{8}{\micro\meter}$ as shot-wise uncertainties for $d_x$ and $d_y$ (see Fig.~\ref{fig:poserrorarray} in supplemental material).
Calculating the error for rotation measurements yields $\sigma_\Omega=\Omega\times\sigma_{d_x}/d_x\approx \SI{0.8}{\milli\radian\per\second}$, for $d_x=\SI{1}{\milli\meter}$ and $\Omega\approx\SI{0.1}{\radian\per\second}$, and for angular acceleration measurements we find $\sigma_{\dot{\Omega}} = \dot\Omega\times\sigma_{d_y}/d_y\approx\SI{0.4}{\radian\per\second\squared}$ with $d_y=\SI{0.4}{\milli\meter}$ and $\dot\Omega=\SI{20}{\radian\per\second\squared}$.
Assuming the positional uncertainties in $d_y$ are solely velocity errors $\sigma_{v_y}=\sigma_{d_y}/t_\mathrm{ff}$, where $t_\mathrm{ff}=\SI{22}{\milli\second}$ is the free fall time after the interferometer before we take absorption images with the camera, we can calculate the rotation sensitivity of this parasitic effect as $\delta\Omega^\mathrm{Sagnac}_i=\delta\phi_{i}^\mathrm{SN}/(2k_\mathrm{eff}T^2\sigma_{v_y})$.
This worst case estimate of $\sigma_{v_y}$ yields $\delta\Omega^\mathrm{Sagnac}_j\approx\SI{224}{\milli\radian\per\second}$, being on the order of our estimated sensitivity limit towards rotation rates. 
\\
Conventional atom interferometers are commonly limited to low rotation rates due to a diminishing contrast, $C \propto \exp[-(\sigma_v k_\mathrm{eff}T^2 \Omega)^2]$, scaling with the expansion velocity $\sigma_v$, as the two branches of the interferometer no longer close in phase space~\cite{Roura14NJP}. Thus, sensors based on thermal ensembles are limited to rotation rates $\Omega \sim \SI{100}{\milli\radian\per\second}$ and require rotation compensation~\cite{decastanet2024atom,sato2024closed}.
The quantum sensor demonstrated here is highly sensitive at even large rotation rates, due to the favorable $\propto \Omega^2$ scaling and the low expansion velocities of the BECs. 
For the comparably high angular velocities of $\Omega\sim\SI{100}{\milli\radian\per\second}$ considered in the presented measurements we achieve a contrast of $\sim 0.6$. We estimate it to drop below $0.4$ for rotation rates higher than $\SI{500}{\milli\radian\per\second}$, yielding a fundamental sensitivity of $\delta\Omega_{j}^\mathrm{SN} = \SI{61}{\milli\radian\per\second}$ for this rotation rate at ensemble temperatures of \SI{9}{\nano\kelvin}. Further collimation of the BECs can push this limit orders of magnitude further~\cite{Herbst24DKC}. 
\\
\textbf{Conclusion and outlook -}
We have demonstrated 
a novel approach to multi-axis inertial sensing by using a 2D BEC array as input for MZAI to realise a measurement of up to 3 inertial quantities simultaneously. 
By differential readout of the simultaneous interferometers at mid-fringe row- and column wise, we were able to measure the linear acceleration, angular velocity, and angular acceleration of the reference mirror.
The method's overall sensitivity to inertial effects is solely limited by technical constraints, such as vibration isolation levels, short baselines, and the orientation control of the sensitive axis. 
The interrogation time in the present experiment is bounded to a few milliseconds as the BECs fall out of the horizontal atom optics light field due to gravity.
We furthermore envision improved sensitivities using guided matter waves~\cite{McDonald14EPL,Boshier_gyro}, or large momentum transfer for small evolution times $T$ as shown for Floquet states~\cite{rodzinka2024floquet,Wilkason2022floquet}. In addition hybridization with classical accelerometers or gyroscopes would allow for post correction of vibration noise sources and lift the ambiguity of acceleration gradient and angular velocity measurements.
Moreover, our method would greatly benefit from operation in long-baseline interferometers~\cite{overstreet_observation_2022,Schlippert2020,zhou_toward_2022}, where interrogation times on the order of seconds are accessible, thus enabling differential measurements of Earth's rotation. Ultimately, if such a device is embarked on a satellite, new avenues in space quantum gravimetry, Earth observation or inertial navigation will be open benefiting a wide spectrum of end users~\cite{leveque2022,ahlers_ste-quest_2022}.  
It is worth pointing out that the continuous scalability of array spacings will enable detailed studies of systematic effects and scale factor calibrations.
\\
To fully characterize the motion of a moving body, an inertial measurement unit (IMU) must capture the acceleration and rotation along three perpendicular directions. We envision applications of our method in compact inertial sensors for dynamic environments~\cite{cheiney_navigation-compatible_2018,decastanet2024atom} by extending the array in the z-axis. A $2\times2\times2$ tensor would enable usage as a 6D-quantum IMU, by either subsequently performing 1D-acceleration measurements in $x$, $y$ and $z$-direction, or by simultaneously applying the atom optics light field from all directions.
\\
Finally, we anticipate exciting applications of our method for in-situ characterization of electromagnetic fields. 
Since each BEC effectively measures a local projection of the wave vector, it becomes possible to systematically study wave front aberrations with high spatial resolution~\cite{bade2018observation}.
This way PIXL acts as an atomic wave front sensor.
Likewise, by observing Rabi oscillations and step-wise translating the array, a full three-dimensional reconstruction of the light field may be realized.
\\
\textbf{Acknowledgements -} We thank G. M\"uller, B. Piest, C. Schubert and T. H. Nguyen for valuable discussions. This work is funded by the
Federal Ministry of Education and Research (BMBF)
through the funding program Photonics Research Germany under contract number 13N14875
and
the Deutsche Forschungsgemeinschaft (DFG, German Research Foundation): Project-ID 274200144 - SFB 1227 DQ-mat
(projects A05 and B07), Project-ID 434617780 - SFB 1464 TerraQ (project
A02), through the QuantERA 2021 co-funded project No. 499225223 (SQUEIS) and Germany’s Excellence Strategy - EXC-2123 QuantumFrontiers
- Project-ID 390837967
and
the German Space Agency at the German Aerospace Center (Deutsche Raumfahrtagentur im Deutschen Zentrum f\"ur Luft- und Raumfahrt, DLR) with funds provided by the German Federal Ministry of Economic Affairs and Climate Action due to an enactment of the German Bundestag under Grants No. 50WM2263A (CARIOQA-GE), No. DLR 50WM2041
(PRIMUS-IV) and No. 50WM2253A (AI-Quadrat).

\bibliography{_main}

\newpage
\cleardoublepage
\thispagestyle{empty}
\newpage
\setcounter{figure}{0}
\setcounter{equation}{0}
\setcounter{table}{0}
\makeatletter
\renewcommand{\theequation}{E\arabic{equation}}
\renewcommand{\thefigure}{A\arabic{figure}}
\renewcommand{\thetable}{T\arabic{figure}}
\section{Supplemental material: Multi-axis inertial sensing with 2D arrays of matter waves}

\subsection{Optical dipole trap}
Key feature of the experimental setup is the ODT setup.
The BEC array is created in a sectional plane of both ODT beams in x/y-direction (Fig. 1 in the main text).
The beam is collimated and split into two beam paths (BP1/2 in Fig. 1).
The two beams differ in power and beam waist in the center of the experimental chamber - BP1 has a power of \SI{3.5}{\watt} and a beam waist of \SI{22}{\micro\meter}.
BP2 has \SI{30}{W} of power and a beam waist of \SI{80}{\micro\meter}. One 2D-acousto optical deflector (AOD) (DTSXY-400) per beam path enables control of the position, amplitude and number of modes in the crossed ODT.
The AODs are supplied with signals generated by a software defined radio (SDR) in the horizontal directions.
The software to drive the SDR is custom made and tailored to the experiment. By applying simultaneously multiple frequency ramps to the AODs distinct harmonic wells in the form of a 2D array can be created.
\subsection{Hybrid trap \& Optical evaporation}
Starting point of the production of the 2D BEC array is a precooled ensemble of $\sim 1\times 10^9$ $^{87}$Rb atoms with a temperatur of \SI{560}{\micro\kelvin} in a magnetic trap. Using a microwave (MW) knife~\cite{Kasevich91PRL} the atoms are evaporatively cooled and after \SI{1.05}{s} of evaporation a temperatur of \SI{150}{\micro\kelvin} is reached. At this point the ODT is switched on with maximum power. 
\\
After another \SI{800}{\milli\second} the magnetic field gradient is decreased to \SI{78}{\gauss\per\centi\meter} and further MW knife evaporation yields an ensemble consisting of $1 - 1.5 \times 10^8$ atoms at a temperature of \SI{50}{\micro \kelvin}. 
The atoms transferred into the crossed ODT are now cooled to BEC by optical evaporation. 
This is done by using three modes in each of the two beam paths. 
They have a spatial position of -511.2, -319.5, \SI{-127.8}{\micro\meter} in x and -97, 48.5, \SI{194}{\micro\meter} in y towards the camera taking absorption images in z-direction, coaligned with gravity. After \SI{2.55}{\second} of all optical evaporation the cooling sequence is completed. 
Powers in the ODT beams were exponentially ramped down to \SI{85}{\milli\watt} (BP1) and \SI{437}{\milli \watt} (BP2). The atom number in the resulting $3 \times 3$ 2D BEC array is $\sim$ \num{3.e5}. The BEC grid has spatial  separation of $(d_x,d_y)=(191,48.5)$~\SI{}{\micro\meter} in the x / y - plane. 
\subsection{State preparation \& Optical transport} 
For transferring the ensembles in the $\ket{F=1,m_F=0}$ state a sequence of Rabi-$\pi$ pulses of the MW is used. 
Active magnetic field stabilisation in $z$-direction compensates for earth's magnetic field and lifts the degeneracy of the hyperfine levels with $B_z=$ \SI{3.69}{\gauss}. 
The state preparation has an efficiency of $> 0.95$ and remaining atoms in the magnetic sensitive states are removed by switching on the magnetic field for one second.
\\
\begin{figure}[h]
    \begin{center}
     \resizebox{0.79\columnwidth}{!}{\includegraphics[width=1\textwidth]{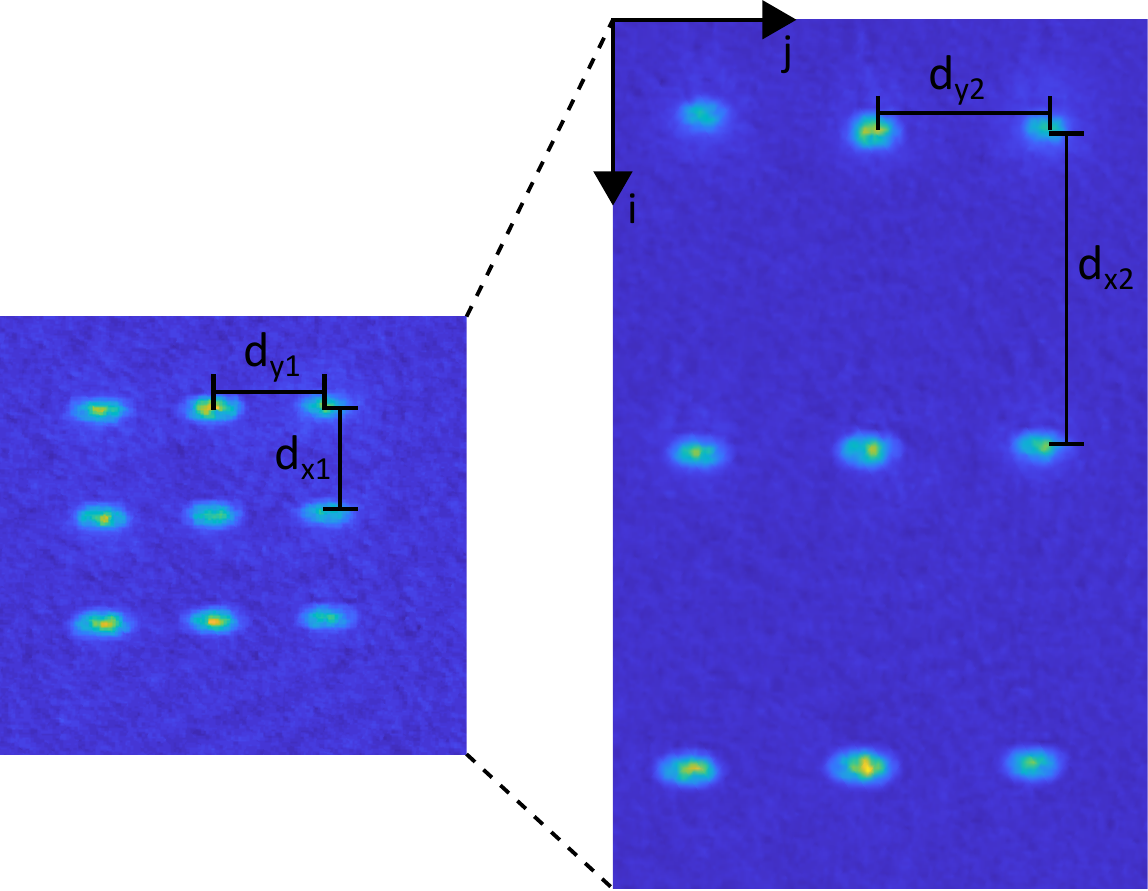}}
    \end{center}
    \caption{\textbf{Absorption images prior to and after the transport ramp.} The distances $d_{x1/y1}$ are increased to $d_{x2/y2}$ after applying the sigmoidal frequency ramps shown in \ref{fig:opttrans}. The Bose-Einstein condensates of the array are labeled using $(i,j)$ for the condensate in the $i$th row and $j$th column.}
    \label{fig:abspic}
\end{figure}
\begin{figure}[t!]
    \begin{center}
     \resizebox{0.99\columnwidth}{!}{\includegraphics[width=1\textwidth]{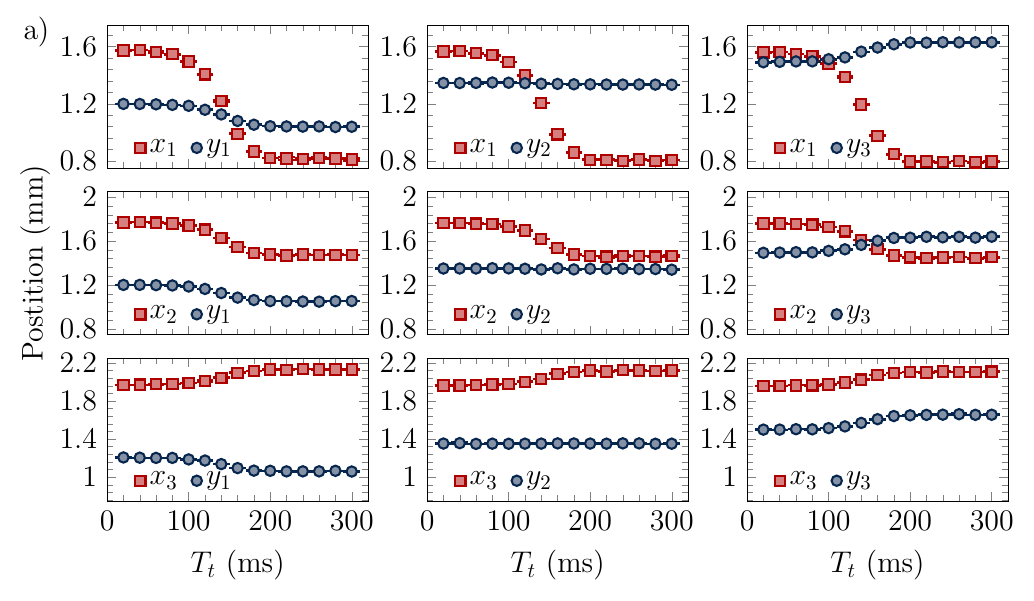}}
     \resizebox{0.99\columnwidth}{!}{\includegraphics[width=1\textwidth]{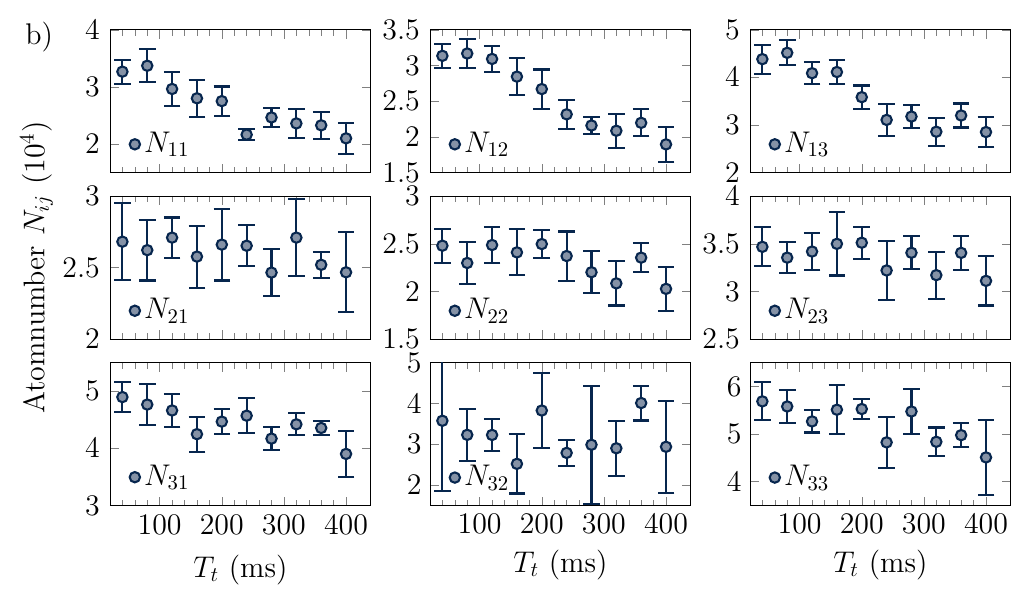}}
     \resizebox{0.99\columnwidth}{!}{\includegraphics[width=1\textwidth]{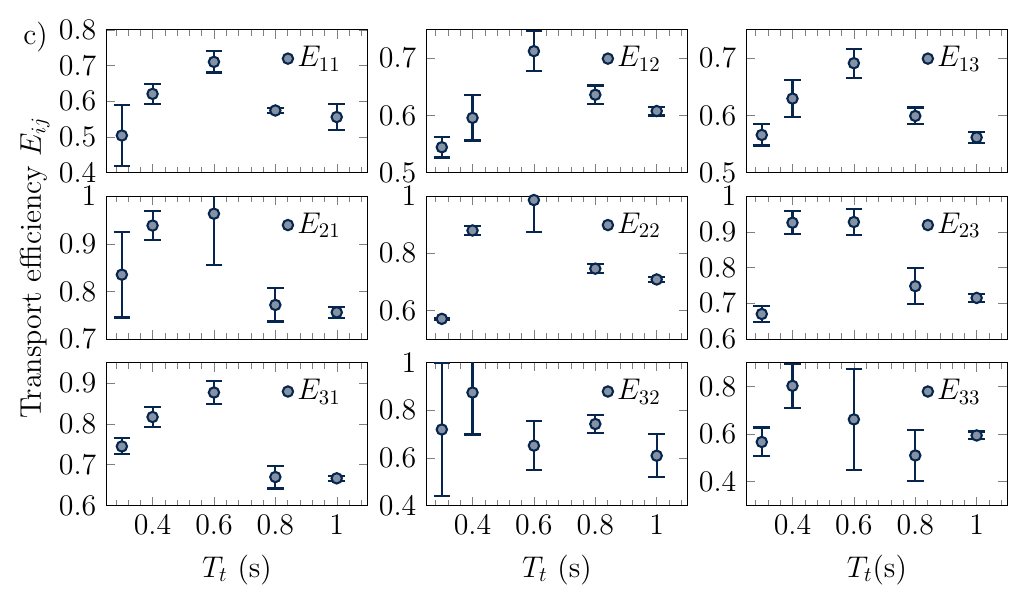}}
    \end{center}
    \caption{\textbf{Characterisation of a transport ramp for the $3\times 3$-array.} \textbf{a)} Trajectories for the 9 BECs moving in $x/y$-direction (red/blue). The resulting grid has a spacing of $(d_x,d_y)=(0.6,0.3)$~mm. \textbf{b)} Atom numbers in the BECs for a \SI{400}{\milli\second} transport ramp. \textbf{c)} Efficiencies of the transport, meaning the atom number at the start of the ramp divided by the atom number at the end of the transport ramp.}
    \label{fig:opttrans}
\end{figure}
For transporting the ensembles, sigmoidal frequency ramps are applied to the AODs \cite{Torrontegui2011}, effectively stretching the array. The final positions of the BECs is set to -660.3, -21.3, ~\SI{617.7}{\micro\meter} in x- and -242.5, 48.5, ~\SI{339.5}{\micro\meter} in y-direction.
In Fig. \ref{fig:abspic} absorption images before and after the transport are shown and in Fig. \ref{fig:opttrans} a) the trajectories of the BECs in the $x/y$-direction are displayed. The resulting $3\times3$-grid has a spacing of $(d_x,d_y)=(0.6,0.3)$~mm, a bit smaller then the array used in the main body. However, the position separation of the BECs in the array is on the order of approximately \SI{1}{\milli\meter} and can be extended to $(d_x,d_y)=(1.0,0.435)$~mm. It is primarily limited by the sectional plane of the two intersecting light fields of the crossed ODT and their Rayleigh range. In Fig. \ref{fig:opttrans} b) the atom number of each BEC is shown for a transportation time of \SI{400}{\milli\second}.
The efficiencies $E_{ij}$ of the transport ramps range from .6 to .95 depending on the transported distance (Fig. \ref{fig:opttrans} c)). 
\\
We observe differing trapping frequencies which can be explained by different amplitudes of the RF-signal supplying the AOD and a tilt in the sectional plane of the two laser beams. The trap frequency measurement yields following values for the trap frequencies in x / y / z [Hz]: 93.7, 100.1, 104.4, 115.1, 122.5, 126.8, 128.9, 136.9, 139.6 / 17.6, 24.8, 28.6, 24.7, 28.6, 30.3, 23.3, 28.6, 30.5 /  176.9, 170.4, 176.2, 149.0, 157.9, 165.3, 138.5, -, - for the BECs $1-9$. For the measurement of the trap frequency in $z$-direction fluorescence detection was used such that not all nine ensembles were sufficiently resolved. This is why ensemble $(3,2)$ and $(3,3)$ are missing. The effective temperatures of the samples ranges from 6 - 9 \SI{}{\nano\kelvin} in $x$, $z$ - direction. The now well separated BEC grid can be used for atom interferometry.
\subsection{Interferometry setup \& Bragg diffraction}
The light pulses to drive the $\pi/2$- and $\pi$-pulses of the interferometer are double-Bragg diffraction pulses \cite{Giese13PRA, küber2016experimental}. We use a frequency doubled Lumibird Keopsys CEFL-KILO MOPA (BL) with output power of \SI{200}{\milli\watt} for interferometry and for referencing the cooling and repump laser that are used for the cooling steps prior to the ODT. The BL is locked to the $^{85}$Rb cooling transition via modulation transfer spectroscopy ~\cite{Noh2011MTS}.
\\
\begin{figure}[
h!]
    \begin{center}   \resizebox{0.99\columnwidth}{!}{\includegraphics[width=1\textwidth]{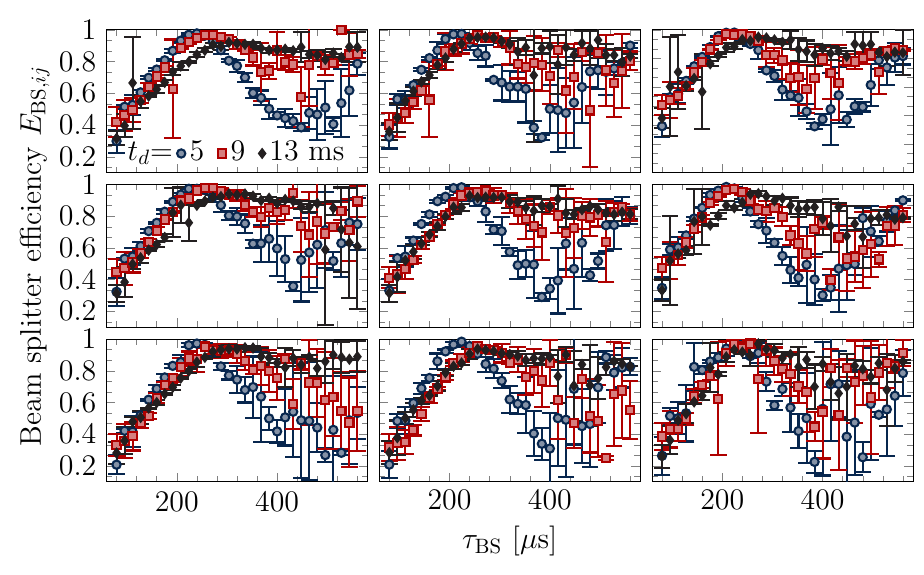}}
    \end{center}
    \caption{\textbf{Rabi oscillations of the double Bragg diffraction.} The oscillations were measured by scanning the pulse duration $\tau_{BS}$ for different pre-tof durations $t_d$. Since the atoms fall through the Bragg beam with a diameter of \SI{8}{\milli\meter} $1/e^2$ the oscillation retardes and the efficiency drops as $t_d$ increases. The array had dimensions of $(d_x,d_y)=(660,290)$~\SI{}{\micro\meter}.} 
    \label{fig:spliteff}
\end{figure}
\begin{figure*}[t!]
    \begin{center}
     \resizebox{1.99\columnwidth}{!}{\includegraphics[width=1\textwidth]{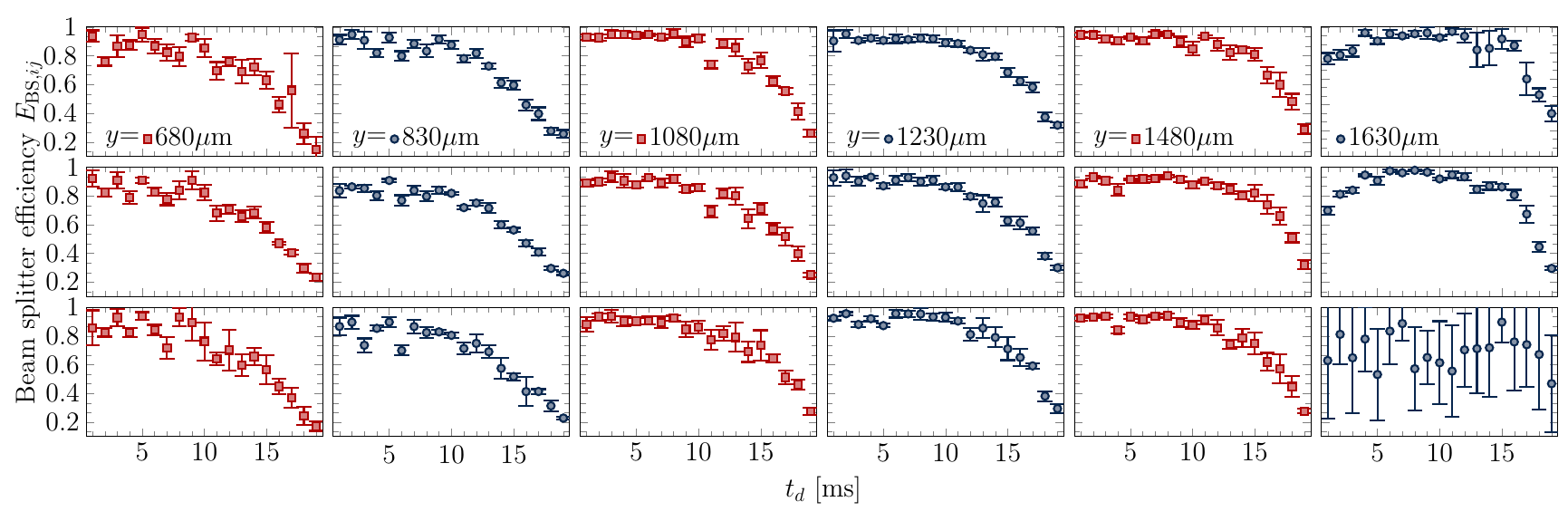}}
    \end{center}
    \caption{\textbf{Pulse efficiency for different y-positions.} By shifting the array by \SI{150}{\micro\meter} and scanning the intitial time-of-flight $t_d$ the splitting efficiency was mapped transversal to the Bragg-beam. The efficiency drops from right to left, meaning that the Bragg-beams center is slightly shifted to the right, regarding the 2D BEC array.} 
    \label{fig:yscanpretof}
\end{figure*}
A total of \SI{120}{\milli\watt} is contributed to the system and first split at a polarisation beam splitter (PBS). Both beams are thereafter diffracted with two AOMs driven at 200 and \SI{200.015}{\mega\hertz} and the first orders are recombined at a second PBS to overlap them and ensure orthogonal polarisations. 
Before tranferring them into a fiber leading to the experiment, they are directed through another AOM for switching. The RF-signal supplied into this AOM has a Gaussian amplitude envelope, for optimal efficiencies of the Bragg-pulses \cite{Parker2016PRA}. A total of \SI{12}{\milli\watt} in both beams is transferred to the experimental chamber.
\\
The beam is collimated to \SI{8}{\milli\meter} $1/e^2$ beam diameter and coaligned with BP1 of the ODT. Before and after passing through the chamber a dichroic mirror is used to separate ODT and Bragg beam. The Bragg beam is passing a $\lambda/4$-plate and is retro reflected at a mirror, rotating the polarity of both back reflected beam components by $\pi/2$. This enables Double-Bragg driven atom interferometry with momentum transfer $\pm \hbar k_\mathrm{eff}$. The beam splitter/mirror pulse has an efficiency of .95/.80 (Fig. \ref{fig:spliteff} depicting Rabi-oscillations of the beam splitter). 
\\
The retro reflection mirror is glued to a 3D-printed mount housing a Thorlabs APF503 piezo (red box in Fig. 1 in the main text). By applying suitable voltage ramps during the interferometry sequence, the mirror can be tilted in x-direction resulting in an effective rotation of the mirror around the z-axis. The mirror is moreover tilted in z-Axis by an angle $\theta_g < 3^\circ$ to induce accelerations $a_{x}=g \times \sin{\theta_g}$, with the gravitational acceleration $g$. The interferometry pulses are applied after transport and another \SI{1}{\second} of wait time, to calm down any exitations induced by the transport. The ODT beams are switched off and the atoms are free falling in z-direction. After an initial time-of-flight (pre-tof) $t_d$ we apply the first interferometry pulse, starting the interferometer sequence. 
\\
Since the laser light used for the Bragg-transitions is a Gaussian profile with a $1/e^2$ diameter of \SI{8}{\milli\meter}, collimated with a telescope before passing the experimental chamber, the intensity of the Gaussian profile can be sampled with the 2D array. This is due to the Rabi-frequency being proportional to $\sqrt{I_1I_2}$ of the two components of the light field. For characterising the beam splitter, two measurements were made: In the first (Fig. \ref{fig:spliteff}) the pulse duration $\tau_{\mathrm{BS}}$ was scanned for different pre-tofs $t_d = $5, 9, 13 \SI{}{\milli\second}, resulting in a retardation and lower amplitude of the Rabi-frequency for longer $t_d$ durations.
The second measurement was taken for optimal pulse duration $\tau_\mathrm{BS}$, but the pre-tof $t_d$ was scanned. The 2D array was shifted for the second run of the experiment by \SI{150}{\micro\meter} in y-direction indicated by the two colors blue/red in Fig. \ref{fig:yscanpretof}.
A rise of efficiency from small to big y-values indicates that the Bragg-beam's center is located at the rightmost column of the measurement or even further to the right. However, the Bragg-beam could not be moved further to the left, due to clipping at the 1'-dichroic mirror separating Bragg- and dipole trap beam.
\subsection{Calibration measurement without induced rotation} 
\begin{figure}[h]
    \begin{center}
     \resizebox{0.99\columnwidth}{!}{\includegraphics[width=1\textwidth]{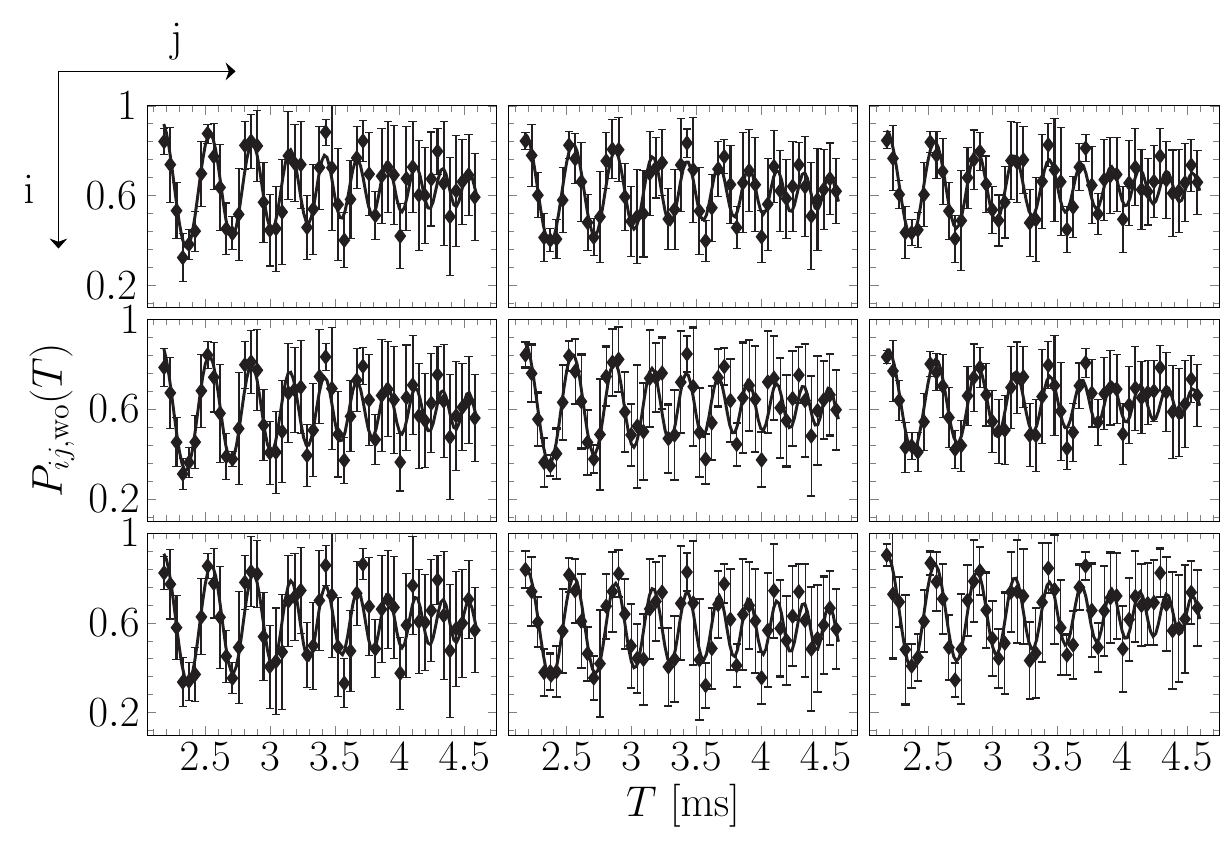}}
    \end{center}
    \caption{\textbf{Callibration measurement without induced rotations.} Fringes obtained for the 9 simultaneous atom interferometers for $T=2.182-4.582$ \SI{}{\milli\second}.} 
    \label{fig:fringeswo}
\end{figure}

In a first step, we perform a calibration measurement without rotating the mirror to identify residual accelerations, e.g., due to a tilt of the sensitive axis with respect to gravity. Note that this is not necessary to extract the linear acceleration from the rotation measurement.
For a non-rotating mirror, the phase shift simplifies to $\phi_{ij}\rvert_{\alpha=0} = 2 k_\mathrm{eff}T^2a_{x,ij}$. 
A fit of the associated fringes $P_{ij}\rvert_{\alpha=0}$ has shown $\phi_\mathrm{0} = 0.4\pi$ and $a_{x,ij}= \{0.1111(7)$, $0.1113(6)$, $0.1109(6)$, $0.1109(7)$, $0.1117(6)$, $0.1111(6)$, $0.1112(7)$, $0.1119(6)$, $0.1107(6)\}$ \SI{}{\meter\per\square\second} for the 9 interferometers top left to bottom right (Fig. \ref{fig:fringeswo}). 
Due to the fact that the Bragg pulses were optimised for a rotating mirror deflecting the beam by $\sim$ \SI{1}{\milli\meter}, the fringe scans with no mirror rotation incorporate an additional frequency $\propto a_{x,ij}/2$ leading to alternating visibility with each oscillation. This is owed to an unfavorable efficiency of the beam splitting pulses, as theoretically confirmed by Ref.~\cite{Li2024PRR}.
Moreover, the differential linear acceleration between two BECs of the grid was measured to be non-zero and can be explained by a position fluctuation of the wave vector's projection onto gravity. Interestingly, this suggests a direct measurement of the Bragg beam's wavefront.

\subsection{Extracting the inertial quantities from the fringe scans}

The signal under consideration, given by the relative population $P_{ij}$ for each BEC, is provided in eq.~(4) of the main text.
For each BEC $(i,j)$, the population offset $P_{ij,0}$ can be determined based on the minimum and maximum population values.
Around this midpoint, the population $P_{ij}$ can be approximated to be linear in the atom interferometer phase.
As detailed in the main text, we focus on values around the mid-fringe, which we define to be $|P_{ij}(T)-P_{ij,0}| \leq 0.11\times C_{ij}$.
The times $T$ where this condition is fulfilled are denoted with $T_\mathrm{mf}$.
At these times, the measured population values can be approximated as
\begin{align}
    P_{ij}(T_\mathrm{mf}) \approx P_{ij,0} + \tfrac{1}{2}\chi_{ij}(T_\mathrm{mf})C_{ij} \left[\phi_{ij}(T_\mathrm{mf}) + \phi_0 + n_{ij}\pi\right],
\end{align}
where $\chi_{ij}(T_\mathrm{mf}) = \mathrm{sgn}[P^\prime_{ij}(T_\mathrm{mf})]$ denotes the sign of the slope of $P_{ij}$. 
$C_{ij}$ is the contrast, $\phi_{ij}$ is the atom interferometer phase (see eq.~(3) in the main text), $\phi_0$ is a phase offset and $n_{ij}\in \mathbb{Z}$ represents the phase ambiguity for each BEC.
From this, the atom interferometer phase is determined as
\begin{align}
    \phi_{ij}(T_\mathrm{mf}) = 
    & \,2\chi_{ij}(T_\mathrm{mf})(P_{ij}(T_\mathrm{mf}) - P_{ij,0})/C_{ij} \notag \\
    & - n_{ij}\pi - \phi_0.
    \label{eq:appendix-phase-from-fringe}
\end{align}

The angular velocity and the angular acceleration can be inferred from the row-wise and column-wise phase differences, respectively (see main text).
Correctly determining the differential phase requires knowing $\chi_{ij}(T_\mathrm{mf})$, which is given by the slope of $P_{ij}(T_\mathrm{mf})$.
Under the dynamic conditions considered here, particularly for large $T$, the slope for an individual measurement point cannot be reliably determined.
While the mean or a fit of $P_{ij}$ provides a good estimate for the slope, noise may still induce phase shifts exceeding $\pi$, effectively altering the slope relative to the mean.
Especially for the angular velocity, which is inferred from the row-wise phase difference for each measurement shot $k$,
\begin{align}
    \phi_{i+1,j}^k(T_\mathrm{mf})-\phi_{i,j}^k(T_\mathrm{mf}) = 4 k_\mathrm{eff} d_x \Omega_M^2(T_\mathrm{mf}) T_\mathrm{mf}^2,
    \label{eq:appendix-diff-phase-angvel}
\end{align}
a natural approach is to take the absolute value of the differential phase, $\Omega_M^2 \propto |\phi_{i+1,j}^k - \phi_{i,j}^k|$.
This ensures that only physically meaningful values are obtained for each measurement shot, as the angular velocity is derived from the square root of the differential phase.
Also, since the phase shift induced by the angular velocity is relatively small, the average slope between rows mostly remains the same (see Fig.~2 in the main text).
However, this approach neglects any differential noise that could turn the phase difference, eq.~\ref{eq:appendix-diff-phase-angvel}, negative.
An alternative approach that largely circumvents this problem is to aggregate the mid-fringe values of all BECs for each individual shot, rather than analyzing them separately.
Assuming that all BECs are mainly affected by a common noise source, their slope directions relative to that of the fit should be identical.
Effectively, for all measurement shots $k$ at times $T_\mathrm{mf}$, the row-wise phase difference is evaluated as
\begin{align}
    \Delta_\mathrm{row}\phi^k(T_\mathrm{mf}) = \left|\left\langle \phi_{i+1,j}^k(T_\mathrm{mf}) -\phi_{i,j}^k(T_\mathrm{mf}) \right\rangle_{i,j} \right|,
    \label{eq:appendix-diff-phase-row}
\end{align}
where $\langle\,\cdot\,\rangle_{i,j}$ denotes the average over all BECs that are in mid-fringe for the respective shot $k$ with the phase given by eq.~\ref{eq:appendix-phase-from-fringe}. 
The uncertainty of $\Delta_\mathrm{row}\phi^k$ is given by the standard deviation of the phase differences per BEC, i.e., of the values inside $\langle\,\cdot\,\rangle_{i,j}$ in eq.~\ref{eq:appendix-diff-phase-row}.
This ensures that differential noise is still accounted for while maintaining a positive overall mean across all measurements $k$, resulting in a physically meaningful outcome.
However, when there are only few BECs in mid-fringe for a respective measurement shot $k$, the confidence in correctly identifying the sign decreases.
In such instances, negative mean values, which arise due to insufficient statistics, are treated as positive values, potentially underestimating the differential phase uncertainty and inflating the overall mean.
Thus, we take the absolute value in eq.~\ref{eq:appendix-diff-phase-row} only when more than one combination of BECs is simultaneously in mid-fringe for a given shot $k$ at time $T_\mathrm{mf}$.
Additionally, if only one combination of BECs is in mid-fringe, we apply the absolute value when the fit at time $T_\mathrm{mf}$ is close to an extremum.
This condition is quantified by the sum of the fringe phases modulo $\pi$, denoted as $\tilde{\phi}_{ij}^k$, i.e., we take the absolute when $|\tilde{\phi}_{i+1,j}^k| + |\tilde{\phi}_{i,j}^k| > \pi/2$.

\begin{figure}[h]
    \centering
    \includegraphics[width=\linewidth]{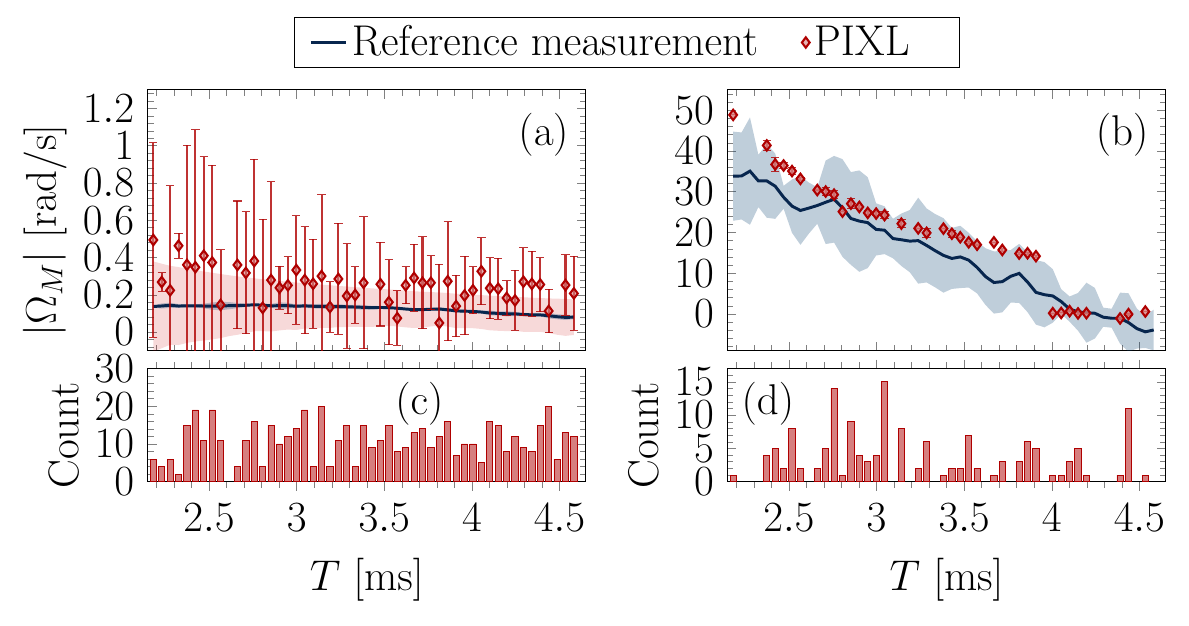}
    \caption{Angular velocity (a) and angular acceleration (b) values including the number of points contributing to the uncertainty, (c) and (d), respectively.} 
    \label{fig:angvel-angacc}
\end{figure}

The angular velocity computed via eqs.~\ref{eq:appendix-diff-phase-row} and~\ref{eq:appendix-diff-phase-angvel}, averaged over all shots $k$, 
\begin{align}
    \Omega_M(T_\mathrm{mf}) = \sqrt{\frac{\left\langle\Delta_\mathrm{row}\phi^k(T_\mathrm{mf})\rvert_{n_{ij}=0} \right\rangle_k }{4 k_\mathrm{eff} d_x T_\mathrm{mf}^2 }}
\end{align}
is shown in Fig.~\ref{fig:angvel-angacc} (a).
The phase offset between rows is assumed to be zero, $\Delta_\mathrm{row} n_{ij} = 0$, since the differential phase induced by the angular velocity is small.
The angular velocity obtained from the differential phases shows good agreement with the one obtained from the reference measurement.
However, as discussed earlier, the mean of most angular velocity values is slightly higher than the reference measurement.
This systematic error is corrected by combining this measurement with the angular acceleration (cf. Fig~\ref{fig:angvel-angacc} (b)), as detailed in the main text (cf. dashed line in Fig.~3 therein).
The number of phase measurements contributing to the angular velocity values is shown in Fig.~\ref{fig:angvel-angacc} (c).
This number varies significantly for different times $T$, leading to irregular uncertainties.

The angular acceleration is obtained analogously from the column-wise phase differences,
\begin{align}
    \phi_{i,j+1}^k(T_\mathrm{mf})-\phi_{i,j}^k(T_\mathrm{mf}) = 2 k_\mathrm{eff} d_y \dot\Omega_M(T_\mathrm{mf}) T_\mathrm{mf}^2.
    \label{eq:appendix-diff-phase-angacc}
\end{align}
Here, the slope direction of an individual mid-fringe measurement is assumed to coincide with that of the fit.
Thus, the column-wise difference of the atom interferometer phases, eq.~\ref{eq:appendix-phase-from-fringe}, is given by
\begin{align}
    \Delta_\mathrm{col}\phi_{i,j}^k(T_\mathrm{mf}) = \phi_{i,j+1}^k(T_\mathrm{mf}) - \phi_{i,j}^k(T_\mathrm{mf}),
    \label{eq:appendix-diff-phase-col}
\end{align}
where the final phase difference is determined by averaging over all measurement shots $k$ and BECs $(i,j)$.
From the scan of the fringes shown in Fig.~2 in the main text, it is evident that the phase obtained from the mid-fringes is additionally offset by $n\pi$, $n\in \mathbb{Z} \setminus \{0\}$, from one column to another.
Given the variation of the angular velocity, we assume a differential phase offset $\Delta_\mathrm{col} n_{ij}=1$.
Thus, we set $n_{ij} = j$.
After $T=\SI{4}{\milli\second}$, the column-wise fringes are in phase again (cf. Fig.~\ref{fig:columnwisephases}) and, thus, we assume $\Delta_\mathrm{col} n_{ij}=0$ for $T_\mathrm{mf}>\SI{4}{\milli\second}$.
The resulting angular acceleration, obtained from the phase differences, eq.~\ref{eq:appendix-diff-phase-col},
\begin{align}
    \dot\Omega_M(T_\mathrm{mf}) = \frac{\left\langle\Delta_\mathrm{row}\phi_{ij}^k(T_\mathrm{mf}) \rvert_{n_{ij}=j}\right\rangle_{k,ij}}{2 k_\mathrm{eff} d_y T_\mathrm{mf}^2},
\end{align}
is shown in Fig.~\ref{fig:angvel-angacc} (b) with the number of measurement points contributing to the signal in Fig.~\ref{fig:angvel-angacc} (d).
It shows good agreement with the reference measurement.
Because of the relatively large phase offset between columns, per measurement shot, there are less mid-fringe values contributing to the angular acceleration value.

\subsection{Theoretical prediction of the atom interferometer phase shift} 
Rotations of the atom interferometer directly couple into the phase shift scaling with the initial kinematics of the atoms. To derive the exact phase shift induced by a certain external rotation, we follow the approach presented in Refs.~\cite{beaufils2023rotation,Struckmann2024,decastanet2024atom}. First, we derive the laser phase $\phi_L$ at a time $t$ with the atoms located at position $(x(t),y(t))$ and velocity $(v_{x}(t),v_{y}(t))$ from geometric considerations. We find
\begin{align}
    \phi_L(t) =& k_\mathrm{eff} \cos (\alpha_I(t) - \alpha_M(t)) [ (x_M - x(t)\cos(\alpha_M(t))\notag \\
    &\quad - (y_m + y(t))\sin(\alpha_M(t)) + d_M]\label{eq:appendix-laser-phase}
\end{align}
where $\alpha_I(t)$ and $\alpha_M(t)$ denote the angle of the laser and the mirror with respect to the $x$-axis (Fig. 1 in the main text). The mirror is assumed to have a thickness of $d_M$ with its center of rotation given by $(x_M, y_M)$. 
In the following, we will assume that the laser is aligned with the $x$-axis, $\alpha_I(t) = 0$.
For small mirror angles the laser phase simplifies to
\begin{align}
    \phi_L = k_\mathrm{eff} [
        & (d_M + x_M - x(t)) + (y_M - y(t)) \alpha_M(t) \notag\\
        & + (x(t) - x_M - \tfrac{d_M}{2})\alpha_M(t)^2 \notag\\
        & + \mathcal{O}(\alpha_M(t)^2)
    ].
\end{align}
The atom interferometric measurement samples this laser phase at times $t=0,T,2T$ where $T$ is the interrogation time. The resulting MZAI phase shift is obtained via
\begin{align}
    \phi =& (\phi_L(0)\rvert_\mathrm{up} + \phi_L(0)\rvert_\mathrm{low}) - 2(\phi_L(T)\rvert_\mathrm{up} + \phi_L(T)\rvert_\mathrm{low}) \notag\\ 
    &+ (\phi_L(2T)\rvert_\mathrm{up} + \phi_L(2T)\rvert_\mathrm{low}) \label{eq:appendix-ai-phase},
\end{align}
where $\mathrm{up}$ and $\mathrm{low}$ denote the upper and lower trajectory of the MZAI separated by $2\hbar k_\mathrm{eff}$. Assuming only a linear acceleration along $x$ acting on the atoms yields
\begin{align}
    x(t)\rvert_\mathrm{up} &= \tilde{x}(t) + 
    \begin{dcases}
        \frac{\hbar k_\mathrm{eff}}{m}t, \quad 0 \leq t < T, \\
        \frac{\hbar k_\mathrm{eff}}{m}(2T-t), \quad T < t \leq 2T,
    \end{dcases}\label{eq:appendix-traj-up}\\
    x(t)\rvert_\mathrm{low} &= \tilde{x}(t) -
    \begin{dcases}
        \frac{\hbar k_\mathrm{eff}}{m}t, \quad 0 \leq t < T, \\
        \frac{\hbar k_\mathrm{eff}}{m}(t-2T), \quad T < t \leq 2T,
    \end{dcases}\label{eq:appendix-traj-low}
\end{align}
where $\tilde{x}(t) = x_0 + v_{x,0} t + \tfrac{1}{2}a_xt^2$ and $y(t) = y_0 + v_{y,0} t$.
Inserting the laser phase into the atom interferometer phase, eq.~\eqref{eq:appendix-ai-phase}, results in
\begin{align}
    \phi = 2 k_\mathrm{eff} [
        & x(0) - 2x(T) + x(2T) \notag\\
        & - (\hat y(0)\alpha_M(0) - 2\hat y(T)\alpha_M(T) + \hat y(2T)\alpha_M(2T)) \notag\\
        & + \hat x(0)\alpha_M(0)^2 - 2\hat x(T)\alpha_M(T)^2 + \hat x(2T)\alpha_M(2T)^2 \notag\\
        & + \mathcal{O}(\alpha_M(t)^2)
    ].\label{eq:appendix-ai-phase-1}
\end{align}
where $\hat x(t) = x(t) - x_M - \tfrac{d_M}{2}$ and $\hat y(t) = y_M - y(t)$.
This expression can be further simplified by expanding the mirror angle around the temporal center of the atom interferometer, $\alpha_M(t) = \alpha_{M,0} + \Omega_M(t-T) + \tfrac{1}{2} \dot\Omega_M(t-T)^2$.
The first term of eq.~\eqref{eq:appendix-ai-phase-1} gives the usual acceleration signal $x(0) - 2x(T) + x(2T) = a_x T^2$ by inserting the MZAI trajectories, eqs.~\eqref{eq:appendix-traj-up} and ~\eqref{eq:appendix-traj-low}.
The second term results in an angular acceleration dependency using $\alpha_M(0) - 2\alpha_M(T) + \alpha_M(2T) = \dot\Omega_M T^2 + \mathcal{O}(T^3)$.
Similarly, the third expression results in an angular velocity dependency via $\alpha_M(0)^2 - 2\alpha_M(T)^2 + \alpha_M(2T)^2 = 2\Omega^2T^2 + \mathcal{O}(T^3)$.
Thus, the full atom interferometer phase shift is given by
\begin{align}
    \phi = -2 k_\mathrm{eff} T^2 [&
        a_x + 2v_{y,0}\Omega_M + (y_0 - y_M + v_{y,0}T)\dot\Omega_M \notag\\
        &- 2(x_0 - x_M - \tfrac{d_M}{2} + v_{x,0}T)\Omega_M^2 \notag\\ 
        &+ \mathcal{O}(T^2)
        ].\label{eq:appendix-ai-phase-full}
\end{align}
From the phase shift it is clear, that the linear acceleration $a_x$ induces a common position and velocity independent phase shift. The phase shift induced by the mirror's angular velocity, $\Omega_M$, scales with the initial velocity in $y$-direction and its square is also scaling with the BEC's initial $x$-position (relative to the mirror's center of rotation). The angular acceleration induces a phase shift scaling with the BEC's initial $y$-position. 
Thus, all three inertial observables - $a_x$, $\Omega_M$ and $\dot\Omega_M$ - can be obtained from a single measurement by exploiting the correlations between at least three BECs of different $x$ and $y$ positions.
\subsection{Position stability of the array} 
\begin{figure}[h]
    \begin{center}
     \resizebox{.99\columnwidth}{!}{\includegraphics[width=1\textwidth]{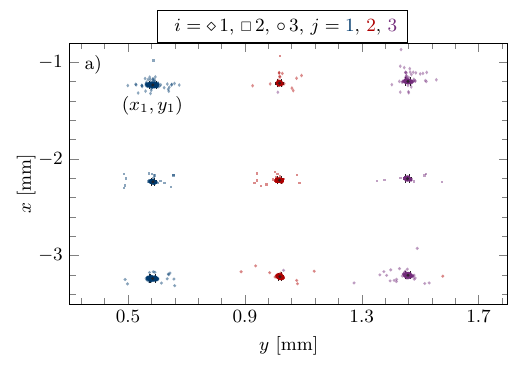}}
     \resizebox{.99\columnwidth}{!}{\includegraphics[width=1\textwidth]{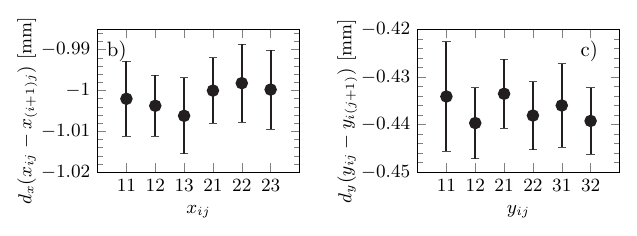}}
    \end{center}
    \caption{\textbf{Positions of the interferometers.} a) In total 4590 data points show the $(x,y)$ positions of the $\ket{0\hbar k_{\mathrm{eff}}}$ state ensembles of each interferometer after the final beam splitter pulse. The points derivating from the densely populated regions are either dropouts, or measurements with vanishingly small residual atom number in the $\ket{0\hbar k_{\mathrm{eff}}}$ state. The error bars show the standard deviation of the data, obtained by removing points further then \SI{70}{\micro\meter} from the mean. b) and c) relative distances between the ensembles of the interferometers row and column wise. The difference was calculated shot wise and the mean value and standard deviation are depicted here.} 
    \label{fig:poserrorarray}
\end{figure}

Following eq.~\ref{eq:appendix-ai-phase-full} the phase is depending on the initital position $(x_0,y_0)$ and velocity $(v_{x,0},v_{y,0})$ of the individual ensembles of the array. For estimating the position and velocity error of the single sites of the array, we fitted the $\ket{0\hbar k_{\mathrm{eff}}}$ state ensembles of each interferometer using Gaussian envelopes and extracted its center in $(x,y)$. The result is shown in Fig.~\ref{fig:poserrorarray} a). 
The mean and standard error of the positions then read (584(12), 1234(12)), (585(6), 2236(10)), (584(9), 3236(11)) for the first column, (1019(6), 1216(8)), (1019(8), 2220(10)), (1020(6), 3218(9)) for the second column and (1459(9), 1197(9)), (1457(6), 2203(8)), (1459(8), 3203(12)) for the third column in \SI{}{\micro\meter}. 
\\
The shot wise differences in the distances $d_x$ and $d_y$ between the ensembles of the interferometers are depicted in Fig.~\ref{fig:poserrorarray} b) and c). They reveal an uncertainty of the distances of $\sigma_{d_x} = \SI{8}{\micro\meter}$ and $\sigma_{d_y} = \SI{8}{\micro\meter}$ and are used to estimate the phase error due to position uncertainty in the main text.

\subsection{\label{sec:appendix-phase-estimation} Phase estimation for time varying scale factors }
After incorporating mirror rotations, we expect signals $P_{ij}(T)$ where the phase consists of offset accelerations $a_x$ and time-varying accelerations $\propto \Omega_M^2(T), \dot\Omega_M(T)$ as described by eq. (2) in the main text.
\begin{figure}[h]
    \begin{center}
     \resizebox{0.99\columnwidth}{!}{\includegraphics[width=1\textwidth]{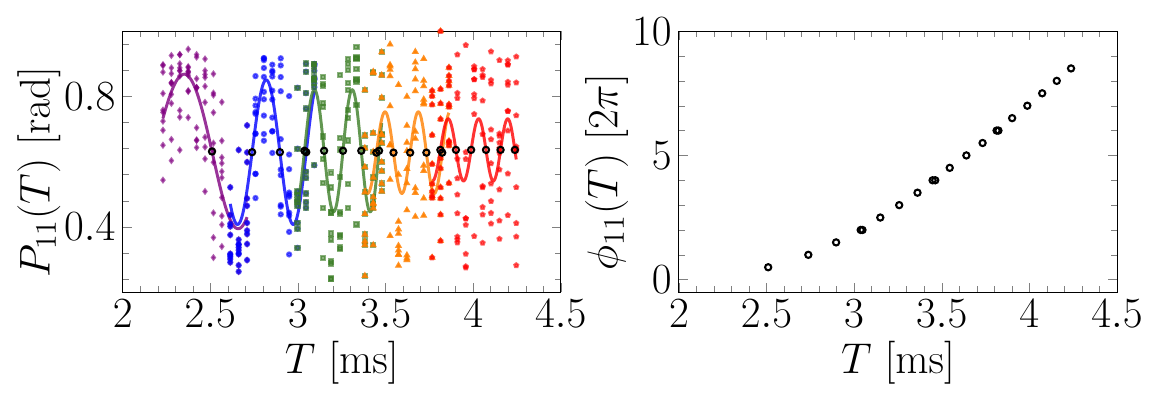}}
    \end{center}
    \caption{\textbf{Principle of extracting the time varying phase $\phi_{ij}(T)$.}
    Mid-fringe points for a sample fringe of the interferometer array, in particular $P_{11}(T)$. The phase is extracted by accumulating one $\pi$ for each of the mid-fringe points (black circles) covered up to the time $T$ of the averaged and interpolated fringes.} 
    \label{fig:fromfringetoscf}
\end{figure}
To extract the time-varying phase from $P_{ij}(T)$, we consider small intervals of $\SI{500}{\micro\second}$, centered around its mean time $t$, where the frequency of the fringe is assumed to be constant. An example of this method is depicted in Fig.~\ref{fig:fromfringetoscf} for the fringe signal $P_{11}(T)$.
\begin{figure}[h]
    \begin{center}
     \resizebox{0.99\columnwidth}{!}{\includegraphics[width=1\textwidth]{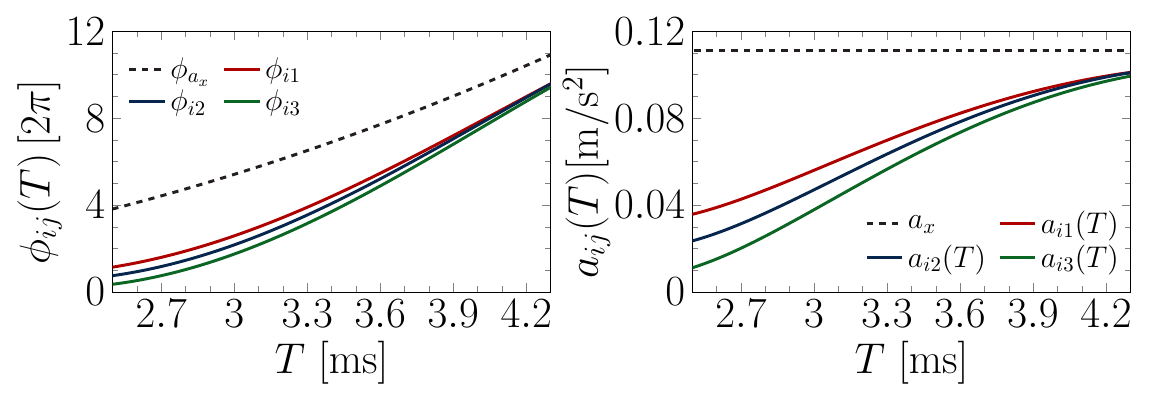}}
    \end{center}
    \caption{\textbf{Column-wise averaged estimation of $\phi_{ij}(T)$  and  $a_{ij}(T)$.} The method shown in \ref{fig:fromfringetoscf} was applied to all nine fringes of the array. By taking the mean of each column the trajectories of the $\phi_{ij}(T)$ were obtained. By using $a_{ij}(T)=\phi_{ij}(T)/(k_\mathrm{eff}T^2)$ the column-wise accelerations $a_{ij}(T)$ were calculated. The dashed black line indicate a quadratic evolution of the phase of the interferometers left and a constant acceleration of $a_x=\SI{0.11}{\meter\per\second\squared}$ on the right.} 
    \label{fig:columnwisephases}
\end{figure}
By moving this interval over the whole dataset, we can accumulate the phase by using $\phi_{ij}(T)=m_{ij}(T)\pi+\phi_{ij,0}$, where $m_{ij}(T)$ is the number of mid-fringe points, covered by the overlapping intervals up to the time $T$ (black circles in Fig.~\ref{fig:fromfringetoscf}) and $\phi_{ij,0}$ is the offset phase.  
\begin{figure}[h]
    \begin{center}
     \resizebox{.99\columnwidth}{!}{\includegraphics[width=1\textwidth]{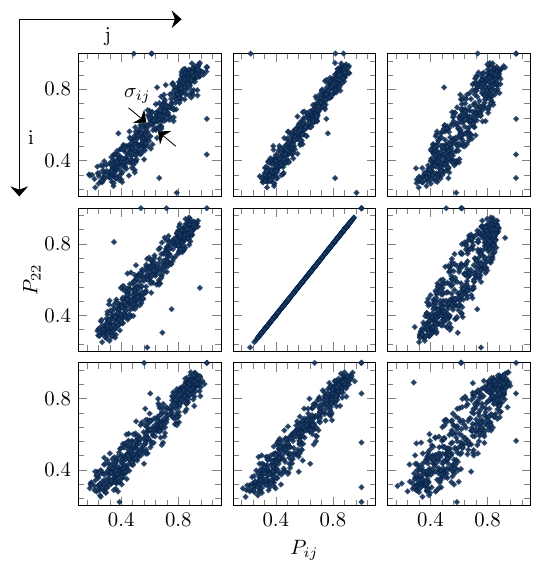}}
    \end{center}
    \caption{\textbf{Parametric plots of the measurement without induced mirror rotation.} Lissajous figures of the measurement without rotation, the $P_{ij}(T)$ are always plotted in respect to $P_{22}(T)$, this is why the center plot shows a line. The width $\sigma_{ij}$ of the Lissajous figures at the mid-fringe position is a measure of the residual noise in the interferometers. We took the data without rotating reference mirror, since the phase shifts due to the rotational acceleration form non trivial Lissajous figures column-wise.} 
    \label{fig:Correlatedwo}
\end{figure}
The phase offset $\phi_{ij,0}$ can be determined by using the averaged fringes $\overline{P_{ij}}(T)$ (see Fig. 2 in the main text) and 
    $P_{ij,\mathrm{fit}}(T)=P_{ij,0}+1/2C_{ij}\times\sin(m_{ij}(T)\pi+\phi_{ij,0})$, where $P_{ij,0}$ and $C_{ij}$ are chosen to approximately match the fringe's offset and contrast.
By taking the difference $\epsilon(\phi_{ij,0})=\overline{P_{ij}}(T) - P_{ij,\mathrm{fit}}(T)$ and minimizing it for $0\leq\phi_{ij,0}<2\pi$ in $P_{ij,\mathrm{fit}}(T)$ we can estimate the phase offset. 
Due to the very small acceleration signal induced by the angular velocity, it is not visible in the phase using this method (Fig. (2) in the main text). Thus, we assume the phase to be the same along the rows. The column-wise phase and the corresponding acceleration $a_{ij}(T)=\phi_{ij}(T)/(k_\mathrm{eff}T^2)$ is depicted in Fig.~\ref{fig:columnwisephases}, where the acceleration agrees with the retardation of the fringes due to the high angular acceleration.
The acceleration induced by the angular velocity and angular acceleration can be extracted from the mid fringe positions. The acceleration induced by a non-zero projection of the sensitive axis with respect to gravity could in principle also be inferred from the central fringes. However, this requires the additional knowledge of the phase offset $\phi_0$. Here, we can fit the phase offset to the fringes depicted in Fig. 2 in the main text and extract the linear acceleration from the phase by subtracting rotational induced accelerations (Fig. 3 in the main text).

\subsection{Common-mode noise rejection ratio}
For estimating the common-mode noise rejection ratio (CMNRR) $\rho_\mathrm{CM}$ of PIXL towards vibrations acting on the reference mirror, we use the data shown in \ref{fig:fringeswo} and plot the output signals $P_{ij}(T)$ parametrically (cf. Fig.~\ref{fig:Correlatedwo}).
\begin{figure}[h]
    \resizebox{1\columnwidth}{!}{\includegraphics[width=.45\textwidth]{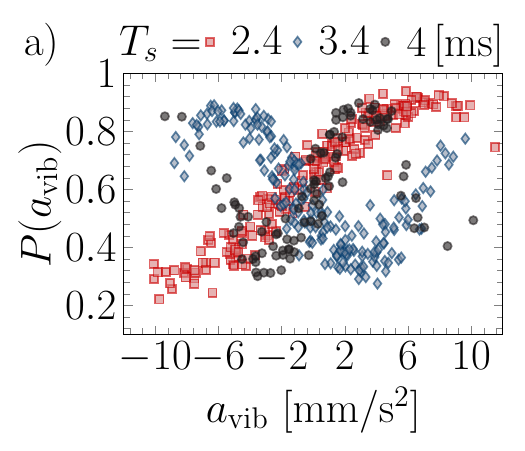}
    \includegraphics[width=.45\textwidth]{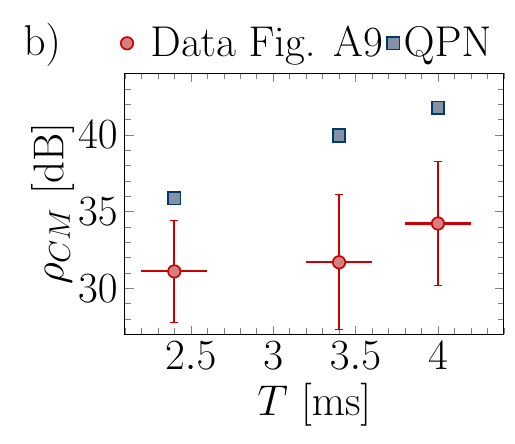}}
    \caption{\textbf{Postcorrected fringes and estimated CMNRR.} a) By tracking the vibrations close to the reference mirror, we were able to reconstruct fringes for a PIXL measurement utilizing a $2\times3$ array at static $T_s=2.4,3.4,\SI{4}{\milli\second}$, averaging the data yields weighted vibrations with Gaussian width $\sigma$ as $a_\mathrm{vib}(T_s)=7.9(7), 6.3(4), \SI{5.6(5)}{\milli\meter\per\second\squared}$. b) Taking the estimated vibration noise for the $T_s$ in a) we calculated the CMNRR for the data given in Fig.~\ref{fig:Correlatedwo} using all mid-fringe datapoints within a \SI{400}{\micro\second} window around the $T_s$ (red horizontal lines). The CMNRR is shown for the average of the 8 datasets of Fig.~\ref{fig:Correlatedwo}, where the error bars are the standard deviation of the 8 mean values. For comparison the CMNRR for the calculated phase uncertainty using the QPN towards accelerations is added to the picture (blue boxes). It was calculated using $\sigma_{ij}(T)=\sqrt{2}/(C\sqrt{N})$ with $C=0.6, N= 10^4$ as in the main text. } 
    \label{fig:postcorrection}
\end{figure}
By calculating the width $\sigma_{ij}$ of the Lissajous figures at mid-fringe we can calculate the residual noise, which we in this case account for pure vibration noise 
\begin{align}
    a_{ij}=\frac{\sigma_{ij}}{2k_\mathrm{eff}T^2}.
\end{align}
For estimating the vibration noise translated to acceleration noise $a_{\mathrm{vib}}$ in the interferometers we use a Nanometrics Titan accelerometer placed close to the reference mirror. This data set was taken after the measurements described in the main body, however the vibration noise level did not significantly change as observed over a period of three months of operation of the accelerometer in the lab. 

By tracking the vibrations on the optical table and using the post correction formalism~\cite{Cheinet2008,Richardson2019, Geiger2011}, we were able to reconstruct fringes for a PIXL measurement utilizing a $2\times3$ array at static $T_s=2.4, 3.4,\SI{4} {\milli\second}$ (cf.~\ref{fig:postcorrection} a) showing the mean over all 6 interferometers). 
The weighted acceleration errors induced by vibrations were estimated to be $a_\mathrm{vib}(T_s)=7.9(7), 6.3(4),\SI{5.6(5)}{\milli\meter\per\second\squared}$ for the times $T_s$, respectively.
Using these measurements we can deduce the minimum CMNRR, given that the residual noise in Fig.~\ref{fig:Correlatedwo} is purely originating from vibrations by~\cite{Bonnin13PRA}
\begin{align}
    \rho_{\mathrm{CM}}=20 \mathrm{log}\left(\frac{a_{\mathrm{vib}}(T_s)}{a_{ij}(T_s)}\right).
\end{align}
We use the mid-fringe points of the Lissajous figures shown in Fig.~\ref{fig:Correlatedwo}
centered around the $T_s$ with interval length of $\sim\SI{400}{\micro\second}$ to obtain sufficient statistics to calculate the average widths $\sigma_{ij}$ for each correlation.
As shown in Fig.~\ref{fig:postcorrection} b) the rejection ratio increases from $\sim\SI{31(3)}{\decibel}$ at $\SI{2.4}{\milli\second}$, to $\sim\SI{34(4)}{\decibel}$ at $\SI{4}{\milli\second}$, corresponding to suppression factors between 35 and 51.

\end{document}